\documentclass[epsfig,useAMS,usenatbib]{mn2e}
 
\input{epsf}

\usepackage {graphics}
\usepackage{graphicx}
\usepackage {layout}

\def\gsim{ \lower .75ex \hbox{$\sim$} \llap{\raise .27ex \hbox{$>$}} }
\def\lsim{ \lower .75ex \hbox{$\sim$} \llap{\raise .27ex \hbox{$<$}} }

\title[Angular momentum transport and disk morphology in SPH simulations
of galaxy formation] {Angular momentum transport and disk morphology in 
SPH simulations of galaxy formation}

\author[Kaufmann et al.]
{Tobias Kaufmann $^1$ \thanks{E-mail: tkaufmann@physik.unizh.ch},
 Lucio Mayer $^{1,2}$, James Wadsley $^3$, Joachim Stadel $^1$\newauthor
{and Ben Moore $^1$  }
\\$^1$ Institute for Theoretical Physics, University of Z\"urich, CH-8057 Z\"urich, Switzerland
\\$^2$ Institute of Astronomy, ETH Z\"urich-H\"onggerberg, CH-8093  Z\"urich, Switzerland
\\$^3$ Department of Physics \& Astronomy, McMaster University, 1280 Main St.
West, Hamilton ON L8S 4M1 Canada}

\begin{document}

\date{Accepted year  Month day. Received year Month day; in original form year Month day}

\pagerange{\pageref{firstpage}--\pageref{lastpage}} \pubyear{} 

\maketitle

\label{firstpage}

\begin{abstract}

We perform controlled N-Body/SPH simulations of disk galaxy formation by
cooling a rotating gaseous mass distribution inside equilibrium cuspy spherical and 
triaxial dark matter halos. We systematically study the angular momentum
transport and the disk morphology as we increase the number of dark matter 
and gas particles from $10^4$ to $10^6$, and decrease the gravitational 
softening from 2 kpc to 50 parsecs.
The angular momentum transport, disk morphology and radial profiles depend 
sensitively on force and mass resolution. At low resolution, similar to 
that used in most current cosmological simulations, the cold 
gas component has lost half of its initial angular momentum via
different mechanisms.  The angular momentum 
is transferred primarily to the hot
halo component, by resolution-dependent hydrodynamical and gravitational 
torques, the latter arising from asymmetries in the mass distribution. In 
addition, disk-particles can lose angular momentum while they are still 
in the hot phase by artificial viscosity. In the central disk, particles can 
transfer away over 99\% of their initial angular momentum due to spiral 
structure and/or the presence of a central bar. The strength of this transport
also depends on force and mass resolution - large softening will suppress the
bar instability, low mass resolution enhances the spiral structure. 
This complex interplay between resolution and angular momentum transfer highlights the complexity of simulations of galaxy formation even in isolated haloes.
With $10^6$ gas and dark 
matter particles, disk particles lose only 10-20\% of their original angular momentum, yet we are unable to produce pure exponential profiles due to the steep density peak of baryons within the central kpc.
We speculate that the central luminosity excess observed in many Sc-Sd 
galaxies may be due to star-formation in gas that has been transported to the
central regions by spiral patterns.
\end{abstract}


\section{Introduction}

Numerical simulations have become the most powerful tool
to explore the complex problem of galaxy formation. Within
the current paradigm of galaxy formation realistic rotationally supported 
disks can form if the gas retains the angular momentum gained
from tidal torques as it cools 
within a cold dark matter
(CDM) halo (White \& Rees 1978; Mo, Mao \& White 1998). However, up to now SPH simulations of galaxy
formation in a cosmological context achieved limited success in supporting
this scenario. The simulated disks were found to be too small and too
centrally concentrated compared to those of observed galaxies, apparently
because the build-up of the galaxy is dominated by cold, dense lumps that
lose too much angular momentum by dynamical friction (Navarro \& Benz 1991; Navarro \& Steinmetz 1997). 
The mismatch in the disk size has been reduced
from the order of magnitude difference of several years ago to a factor
of order two in the most recent simulations thanks to both higher resolution
and the inclusion of energetic feedback mechanisms which partially suppress
the cold lumps (Governato et al. 2004; Sommer-Larsen, G{\"o}tz \& Portinari 2003; Thacker \& Couchman 2000; Abadi et al. 2003; Robertson et al. 2004). Even in the latest simulations the internal structure of galaxies is still barely
resolved (Abadi et al. 2003). Therefore it is unclear how robust these results are, especially since detailed convergence
studies such as those existing in other areas of cosmology and astrophysics  (e.g. Power et al. 2003 for dark matter halos) are still lacking.

Observationally, many galaxies are single component disks with a nearly exponential surface density profile. Reproducing
these galaxies is one of the biggest remaining challenges for the
CDM paradigm. It is not clear if the inability to simulate a pure disk
galaxy with the correct scale length is due to a fundamental flaw in 
the CDM model, or due to missing ingredients in the cosmological simulations, 
or perhaps due to numerical effects such as resolution.
It is claimed that low resolution simulations
$(N\sim10^{4})$ suffer from excessive artificial viscosity that
can cause spurious angular momentum losses 
(Thacker et al. 2000; Sommer-Larsen, Gelato \& Vedel 1999). Moreover, for
stellar disks embedded in poorly resolved halos, two-body-effects will tend to increase the contribution of random kinetic energy relative to that of rotational energy
by heating stars out of the disk plane, producing artificially hot stellar
distributions rather than cold disks (Governato et al. 2004), 
and would induce spurious heating of the gas component as well (Steinmetz \& White 1997).

On the other hand central dense stellar nuclei have been seen in several late-type spiral galaxies, for example, in M33 (Regan \& Vogel 1994), but also in other late-type spirals (Carollo et al. 1998, Carollo 1999), and the  semi-analytical models of  Van den Bosch (2001) produced disk galaxies with surface density profiles which slopes were steeper  near the centre than an exponential.

In this paper we investigate various numerical effects that can potentially
compromise the results of current simulations of galaxy formation. We will
focus in particular on the evolution of the angular momentum of forming disks
and on how that depends on mass and force resolution as well as on other 
parameters of a simulation, including how the initial conditions are set up.
Our study considerably extends a similar one recently carried out by Okamoto et al. (2003).

For this investigation we will use N-Body+SPH simulations having between $10^4$ and $6\times 10^6$ dark
matter and SPH particles. We will also explore the effects
of force resolution at a fixed mass resolution.  The simulations will follow the
formation of a large galaxy with mass comparable to the Milky Way and of a smaller
galaxy with dark and baryonic mass comparable to that of the Local Group spiral M33.
The central disks form by the cooling of gas inside
equilibrium dark matter halos in which the gas is set to
initially rotate in 
accordance with expectations from cosmological simulations, either by construction
or set by mergers between equilibrium non-rotating haloes. 

Cosmological simulations
suggest that the large disks of spiral galaxies form mainly from
the smooth accretion of gas after the last major merger
(Sommer-Larsen, G{\"o}tz \& Portinari 2003; Abadi et al. 2003; 
Governato et al. 2004) and heating by cold dark matter substructure is not
expected to affect disk properties dramatically (Font et al. 2001).
Therefore, although our simulations do not follow the hierarchical growth
of structure in a full cosmological context, they can model with unprecedented
detail the  gas accretion phase which characterises the main
epoch of disk formation. The controlled experiments presented in this paper are thus
complimentary to cosmological simulations, allowing us to directly explore
not only the consequences of varying the resolution, but also the effect of different recipes of star 
formation and different treatments of the gas thermodynamics.

The outline of the paper is as follows. In section 2 we present
the modelling of our initial conditions, and the treatment of 
cooling and star formation. In section 3 we give an overview of  
the results of the simulations and discuss the resolution dependence 
of our results, focusing in particular on the analysis of the 
angular momentum transport. We  discuss the dependence of disk formation on star formation (section 4) and on different
ways of setting up the initial conditions (section 5), we then summarise and conclude in section 6.

\section{Initial Conditions}

Our initial conditions comprise an equilibrium NFW galaxy-sized halo with an embedded
spinning gaseous mass distribution in hydrostatic equilibrium, using models having from 
250 dark matter \& 30,000 gas particles up to 1 million dark matter \&  gas particles. 
We do not include any heating source such as 
the cosmic UV background or feedback from supernovae. In most of the simulations
we enforce a lower limit for the temperature that the gas can reach via radiative
cooling. This can be seen as a crude way to mimic the effect of the
missing heating sources (see also section \ref{sub:tempfloor} and Barnes 2002).

The halo models are built as in Kazantzidis, Magorrian \& Moore (2004), and hence they
include a self-consistent description of the velocity distribution function.
In order to initialise the dark halo we start by choosing
the value of its circular velocity at the
virial radius, $v_{200}$, which, for an assumed cosmology (hereafter
$\Omega_{0}=0.3,\Omega_{\Lambda,0}=0.7,H_{0}=70$ kms$^{-1}$ Mpc$^{-1}$) automatically
determines the virial mass, $M_{200}$, and virial radius, $r_{200}$,
of the halo (Mo, Mao \& White 1998);  $r_{200}$ is the radius corresponding to a density of 200 times the critical density. For the main model used in this work
we choose $v_{200}=140$ km/s,
which corresponds to $r_{200}=200$ kpc and therefore and $M_{200}=9.14\times10^{11}M_{\odot}$. Haloes have
NFW density profiles (Navarro, Frenk \& White 1996) characterised by a halo concentration 
$c$, defined as $c=\frac{r_{200}}{r_{c}},$
where $r_{c}$ is the halo scale radius. For our main model we adopt $c = 8$, which corresponds to $c_{vir}=\frac{r_{vir}}{r_{c}}\approx 11$, where $r_{vir}$ is the actual virial radius in the assumed cosmology. This is at the low end of the values of the concentration for halos with masses in the 
range $5 \times 10^{11}-10^{12} M_{\odot}$ (Bullock et al. 2001a), but it is very similar to the concentration parameter for the best fit halo model of the Milky Way found by Klypin, Zhao \& Somerville (2002), which satisfies a variety of
observational constraints.

The gaseous halo of our fiducial model is constructed in the same way and contributes 
$9.09\%$ of the total mass of the system. The temperature structure
 of the gaseous halo is calculated
by solving the equation for the hydrostatic balance of an ideal gas inside a dark matter halo, 
as done in Mastropietro et al. (2005). Assuming an spherically symmetric model, the halo temperature 
at a given radius $r$ is determined by the cumulative mass distribution $M(r)$ of 
the dark and gaseous components beyond $r$ and by the density profile $\rho_h(r)$ of the hot gas:
\begin{equation}\label{gashalo}
T(r) = \frac{\mu}{k_B} \frac{1}{\rho_h(r)} \int_{r}^{\infty} \rho_h(r)\frac{GM(r)}{r^2} \,dr \, ,
\end{equation} where $\mu$ is the mean molecular weight, $G$
and $k_B$ are the gravitational and Boltzmann constants.
The hot gas starts
with a specific angular momentum distribution motivated by
the results of cosmological simulations; the initial specific angular momentum
profile is well fit by a power law following $j\propto r^{1.0}$,
whereas the dark matter halo has no angular momentum (Bullock
et al. 2001b proposed  a power law with $j(r)\propto r^{\alpha},\alpha=1.1\pm0.3.$).
In the Milky Way models we used $\lambda=0.038$ for the spin parameter,
defined  as $\lambda=\frac{j_{gas}\left|E\right|^{\frac{1}{2}}}{GM^{\frac{3}{2}}}$,
where $j_{gas}$ is the average specific angular momentum of the gas, $E$ and $M$are the total energy and mass of the halo. 
This definition matches the one commonly  used (e.g. Mo, Mao \& White 1998) under the assumption that
there is no angular momentum transport between the spherical dark
matter halo and the gas. This transport turned out to be small, see section $3.1$.

The mass of dark matter particles varies between $4\times10^{9}M_{\odot}$
in the low resolution runs and $1.66\times10^{5}M_{\odot}$ in the
high resolution runs. For the highest resolution models we used a dark matter halo with
variable particle masses, with the resolution increasing towards the centre of
the system. These shell models comprise: $600,000$ particles distributed in the
inner sphere with $20$ kpc radius, $300,000$ particles in the
next shell out to $100$ kpc, while the rest of the halo
is sampled with only $100,000$ particles. 
With a single-resolution model one would need about
ten million particles to reach a comparable resolution in the central region.
This allows us to resolve the central dark matter cusp to a scale of $\sim 100$ parsecs
in the MW model.
The reliability of these
initial conditions is tested and reported in Zemp (2006).  
The mass of gas particles varies from $3.4\times10^{6}M_{\odot}$ in the low resolution
runs to $2.0\times10^{5}M_{\odot}$ in the high resolution runs. For the gas  we did not use a shell model because smoothing over particles
of different masses in an SPH code can produce spurious results (Kitsionas \& Withworth 2002). The softening lengths for the gas and dark matter particles were chosen to be  $\epsilon=2$ kpc - similar as in a typical cosmological simulation (e.g. Sommer-Larsen, G{\"o}tz \& Portinari 2003; Governato et al. 2004) -  and four times smaller $\epsilon=0.5$  kpc for the hi-res run (this being more comparable to the force resolution in the latest generation of galaxy formation runs, see e.g. Governato et al. 2006.
We provide an overview of the main simulations carried out in Table \ref{cap:Overview-to-all-sim}. In section  \ref{merger_section} we describe an alternative way of building the initial conditions that more
closely resembles what happens in a full cosmological framework.

We also carry out a set of simulations with a less massive halo, comparable to that expected for
the nearby spiral galaxy M33 (Corbelli 2003), such that
$M_{200}=5.3\times10^{11}M_{\odot},$ concentration $c=6.2,$
spin parameter $\lambda=0.105$. Since we want to use this model to produce a low mass spiral such as M33 we
also adopt a lower baryon fraction, $f_{b}=6\%$, which after 5 Gyr of cooling gives a similar cold baryonic
component as observed within this galaxy.
We consider two different initial specific angular momentum profiles for the gas, $j\propto r^{1.0}$  (M33A), as
for the Milky Way model,  and $j\propto r^{0.5}$ (M33B) (we keep the total angular momentum inside the virial radius fixed).
In the ``standard'' M33 model the hot gaseous halo is resolved with $5\times 10^5$ particles of mass $\sim10^5M_\odot$.
We will also present some results from the ``refined'' simulation where we resolve the gas phase down to 
$3000$ $M_{\odot}$ using a splitting of the SPH particles (see Kaufmann et al. 2006 and references therein). 
We adopt a force resolution of $\epsilon=0.25$ and $\epsilon=0.1$ kpc for the standard and 
hi-res run, respectively, see Table \ref {cap:Overview-to-all-sim-M33}. In contrast to the Milky Way runs this model cools a lower fraction of baryons into the final disk, which
allows us to run at a higher resolution since the central disk does not achieve such high densities thus 
relaxing the requirements on the time-step size.

Initial conditions similar to those employed here have been used in Robertson et al. (2004)
and Okamoto et al. (2003), however we will compare simulations that have much higher
force and mass resolutions.
Finally, the semi-analytical work of van den Bosch (2001) 
also uses an initial setup that is quite similar in spirit, in which they follow in one 
dimension the collapse of shells of a rotating gas distribution inside an NFW halo.
This makes an interesting comparison since it is naturally free of numerical effects.

To evolve the models we use the parallel TreeSPH code GASOLINE, which is
described in Wadsley, Stadel \& Quinn (2004). The code is an extension to 
the N-Body gravity code PKDGRAV developed by Stadel (2001). 
It uses an artificial viscosity which is the shear
reduced version (Balsara 1995) of the standard Monaghan (1992) artificial
viscosity. GASOLINE solves the energy equation using the asymmetric form (Wadsley, Stadel \& Quinn 2004, equation 8) and conserves entropy closely. It uses the standard spline form
kernel with compact support for the softening of the gravitational
and SPH quantities. The code includes radiative cooling
for a primordial mixture of helium and (atomic) hydrogen. Because of the lack of molecular
cooling and metals, the efficiency of our cooling functions drops rapidly
below 10,000 K. Nearly all simulations adopt a temperature floor which is
higher than this cut-off (see section 5.1) so 
as to crudely mimic the effect of heating sources such as supernovae
explosions and radiation (e.g. ultraviolet) backgrounds.
The star formation recipes are described in section \ref{Starformation}.

\begin{table*}
\begin{tabular}{|c|c|c|c|c|c|}
\hline 
Name&
\multicolumn{1}{c|}{Resolution {\small }{\footnotesize (dark /gas particles)}}&
{\small softening {[}kpc{]}}&
{\small $T_{cut-off}$ {[}kKelvin{]}}&
{\small art. viscosity}&
{\small star formation}\tabularnewline
\hline 
\hline
LR&
{\small $3\times10^{4}$/ $3\times10^{4}$ }&
$2$&
$30$&
standard&
no\tabularnewline
\hline 
LRLS&
{\small $3\times10^{4}$/ $3\times10^{4}$ }&
$0.5$&
$30$&
standard&
no\tabularnewline
\hline 
LRLD&
{\small $2.5\times10^{2}$/ $3\times10^{4}$ }&
$5$&
$30$&
standard&
no\tabularnewline
\hline 
IR&
{\small $9\times10^{4}$ / $9\times10^{4}$ }&
$2$ &
$30$&
standard&
no\tabularnewline
\hline 
IRLS&
{\small $9\times10^{4}$ / $9\times10^{4}$ }&
$0.5$&
$30$&
standard&
no\tabularnewline
\hline 
IRLSSF&
{\small $9\times10^{4}$ / $9\times10^{4}$ }&
 $0.5$ &
$30$&
standard&
Katz 1992\tabularnewline
\hline 
IRLSSFHE&
{\small $9\times10^{4}$ / $9\times10^{4}$ }&
 $0.5$ &
$30$&
standard&
Katz high efficiency\tabularnewline
\hline 
IRLSNL&
{\small $9\times10^{4}$ / $9\times10^{4}$ }&
$0.5$&
$30$&
no shear reduction&
no\tabularnewline
\hline 
IRLSMD&
{\small $9\times10^{5}$ / $9\times10^{4}$ }&
$0.5$&
$30$&
standard&
no\tabularnewline
\hline 
HR&
{\small $1\times10^{6}$ / $5\times10^{5}$ }&
$2$&
$30$&
standard&
no\tabularnewline
\hline 
HRLS&
{\small $1\times10^{6}$/ $5\times10^{5}$ }&
$0.5$&
$30$&
standard&
no\tabularnewline
\hline
HRSFT&
{\small $1\times10^{6}$/ $5\times10^{5}$ }&
$2$&
$30$&
standard&
Temperature criterion \tabularnewline
\hline
\end{tabular}

\caption{Overview of the numerical parameters of the main simulations of the Milky Way model. Here and in the following, LR, IR and HR indicate the resolution used, LS the use of small softening and SF the use of  star formation. The ``standard'' artificial viscosity refers to the Monaghan artificial viscosity described at the end of section 2.}\label{cap:Overview-to-all-sim}
\end{table*}

\begin{table*}
\begin{tabular}{|c|c|c|c|c|c|}
\hline 
Name&
\multicolumn{1}{c|}{Resolution {\small }{\footnotesize (dark /gas particles)}}&
{\small softening {[}kpc{]}}&
{\small $T_{cut-off}$ {[}kKelvin{]}}&
{\small art. viscosity}&
{\small Angular momentum profile}\tabularnewline
\hline 
\hline 
M33A&
{\small $1.1\times10^{6}$/ $5\times10^{5}$ }&
$0.25$&
$30$&
standard&
$\propto r^{1.0}$\tabularnewline
\hline
M33B&
{\small $1.1\times10^{6}$/ $5\times10^{5}$ }&
$0.25$&
$30$&
standard&
$\propto  r^{0.5}$
\tabularnewline
\hline 
refined M33A&
{\small $2.2\times10^{6}$/ $4\times10^{6}$ }&
$0.1$&
$15$&
standard&
$\propto r^{1.0}$\tabularnewline
\hline
\end{tabular}
\caption{Overview to the numerical parameters of the main simulations of the M33 model.}\label{cap:Overview-to-all-sim-M33}
\end{table*}

\section{Results of the simulations}

We first test the equilibrium of the combined gas and dark matter
halo by evolving it using an adiabatic equation of state. 
We find that the mass inside the 
virial radius varies only by $\sim 0.5\%$ after 5 Gyr of evolution
(e.g. for the IR simulation), the temperature and density profiles of the gas also undergo only minimal
variation. In the following we describe the results of the
runs in which the spinning gas cools radiatively inside the halo and gradually 
settles into a centrifugally
supported central object. We run all the simulations for at least five
billion years.
We give an overview of the different runs in table 
\ref{cap:Results-of-the-sim}. We begin with discussing the numerical convergence of 
the basic properties such as the mass and the angular momentum of the
central cold baryons. We then study the morphological evolution of the central
disk and the dependence of results on the initial conditions.

\begin{table}
{\footnotesize }\begin{tabular}{|c|c|c|c|}
\hline 
Name&
central object&
radius {[}kpc{]}&
\tabularnewline
\hline
\hline
LRLD&
blob plus 2 fragments&
2&
\tabularnewline
\hline 
LR&
thick disk&
4&
\tabularnewline
\hline 
LRLS&
thick disk&
4&
\tabularnewline
\hline 
IR&
disk&
5.6&
\tabularnewline
\hline 
IRLS&
barred disk&
6.3&
\tabularnewline
\hline 
IRLSNL&
barred disk&
5.7&
\tabularnewline
\hline 
IRLSSF&
barred disk&
5.9-6.7&
\tabularnewline
\hline 
IRLSSFHE&
barred disk&
6.5-7&
\tabularnewline
\hline 
IRLSMD&
barred disk&
6&
\tabularnewline
\hline 
HR&
disk, oval&
5.3-6&
\tabularnewline
\hline 
HRLS&
barred disk&
6.9&
\tabularnewline
\hline
HRSFT&
disk&
6.6&
\tabularnewline
\hline
\end{tabular}{\footnotesize  \vspace{0.375cm}}{\footnotesize \par}

\caption{The morphology and size (radial extent of the cold gas particles) of the central disk after 5 Gyr of evolution 
for the different simulations of the Milky Way model.}\label{cap:Results-of-the-sim}
\end{table}

\subsection{Convergence in mass and angular momentum}

\begin{figure}
\includegraphics[%
  scale=0.4]{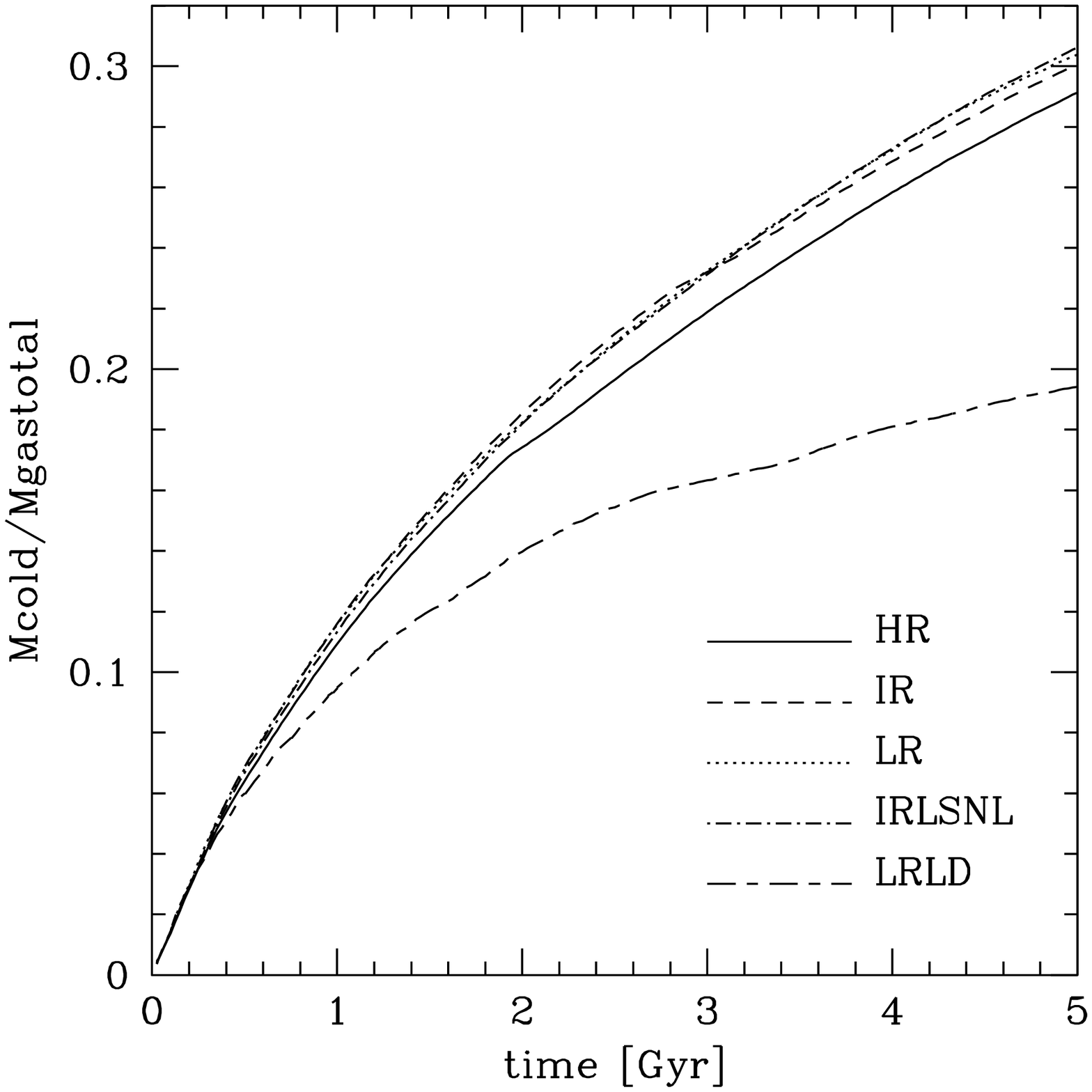} \includegraphics[%
  scale=0.4]{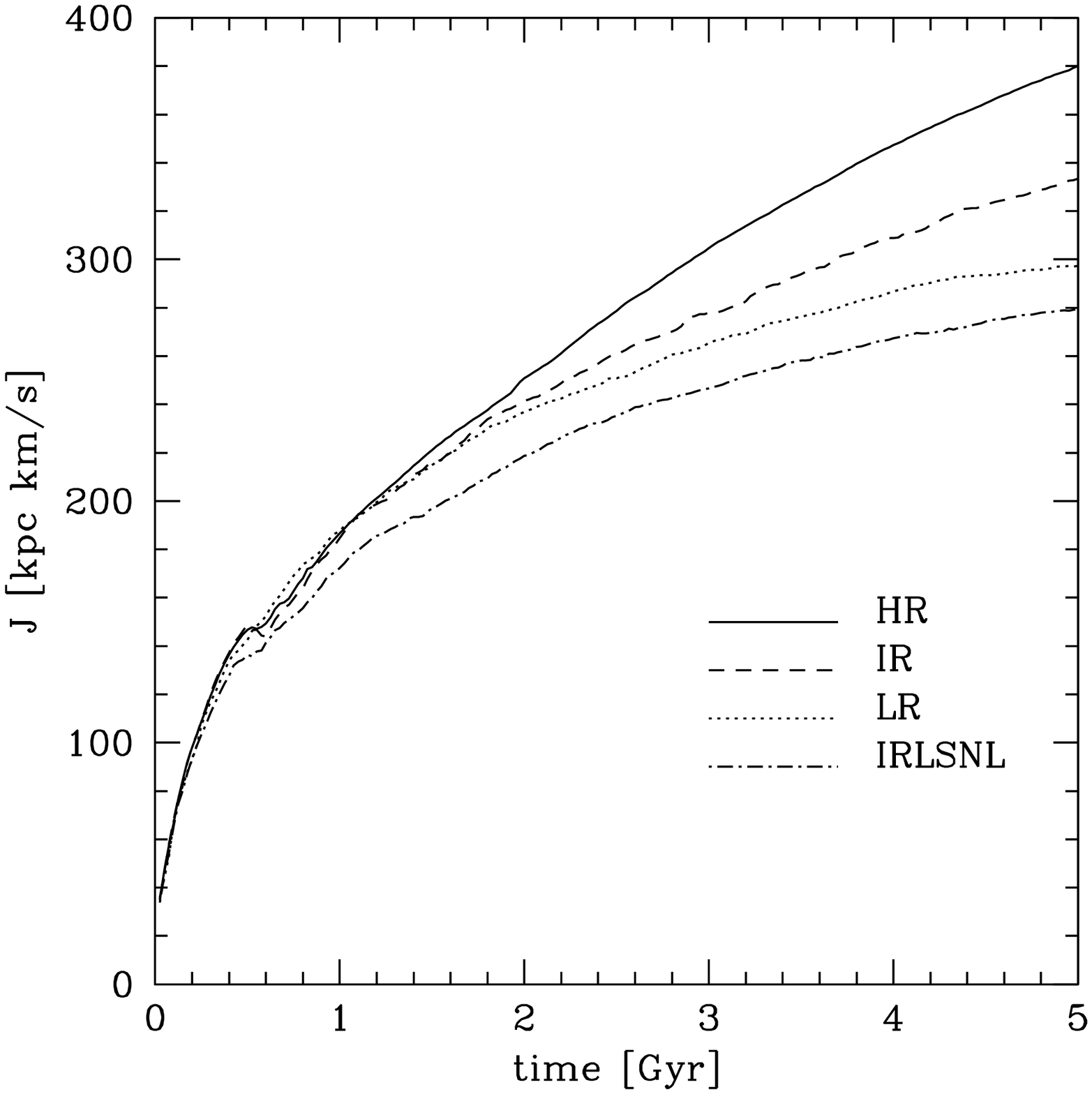}

\caption{Upper panel: Growth of the cold  mass fraction (equivalent to the disk mass fraction) for gas particles that 
were initially inside a sphere of radius $80$ kpc. Lower
panel: Accretion of specific angular momentum of the cold gas disk
for particles within the same sphere. HR is plotted with solid,  IR with 
dashed,  LR with dotted, LRLD with long dashed - short dashed 
and IRLSNL with dashed-dotted lines. We find convergence in 
terms of of evolution of the cold mass fraction but not in terms of evolution of the angular momentum of the cold particles. For the description of the different runs see table \ref{cap:Overview-to-all-sim}. \label{cap:Accretion-of-J/M}}
\end{figure}

The main simulations are done at three different resolutions:
LR, IR, HR. Moreover, in order to compare with the results of Steinmetz \& White (1997),
we re-simulated the LR run with a smaller number of dark particles, such that the
mass of the dark particles exceeds the critical mass given in their
work by a factor of $2$ (run LRLD).  Above this critical mass, two-body encounters are expected to
produce spurious heating that overwhelms radiative cooling, thus suppressing the amount of accreted
mass. We find comparable accretion
of cold gas mass between the runs LR, IR and HR. However, we
confirm qualitatively the results of  Steinmetz \& White (1997) with the LRLD run:
in this case the cold gas mass is reduced by a factor of $1.6$ compared to the other runs
because of spurious heating  see Figure \ref{cap:Accretion-of-J/M}.
(Note that this is slightly lower than that found by those authors because we use a higher cut-off in the cooling function.)

We measured the specific angular
momentum of the central cold particles with $T<80,000$ K. We use this temperature  threshold instead of $30,000$ K (the effective temperature of the disk) to include also those particles that are near the edge of the disk and are about to be accreted. We checked that the results reported below are not strongly affected by the precise way in which we identify cold particles (e.g. $T<50,000$ K or $T<80,000$ K).

\begin{figure*}
\includegraphics[%
  scale=0.614]{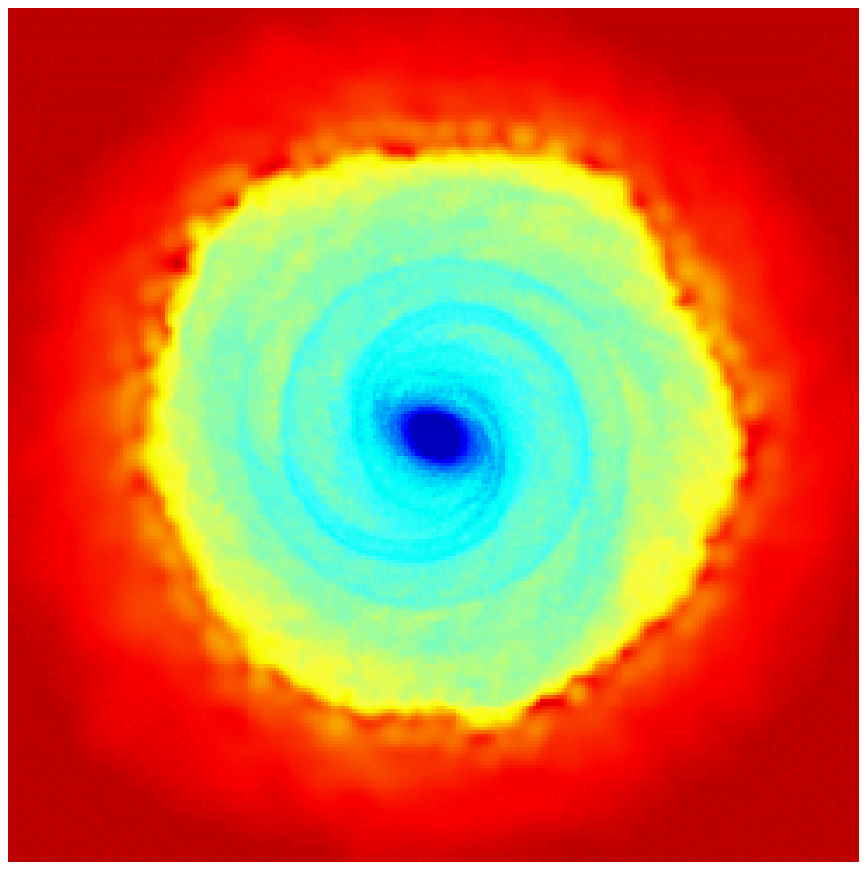}\includegraphics[%
  scale=0.614]{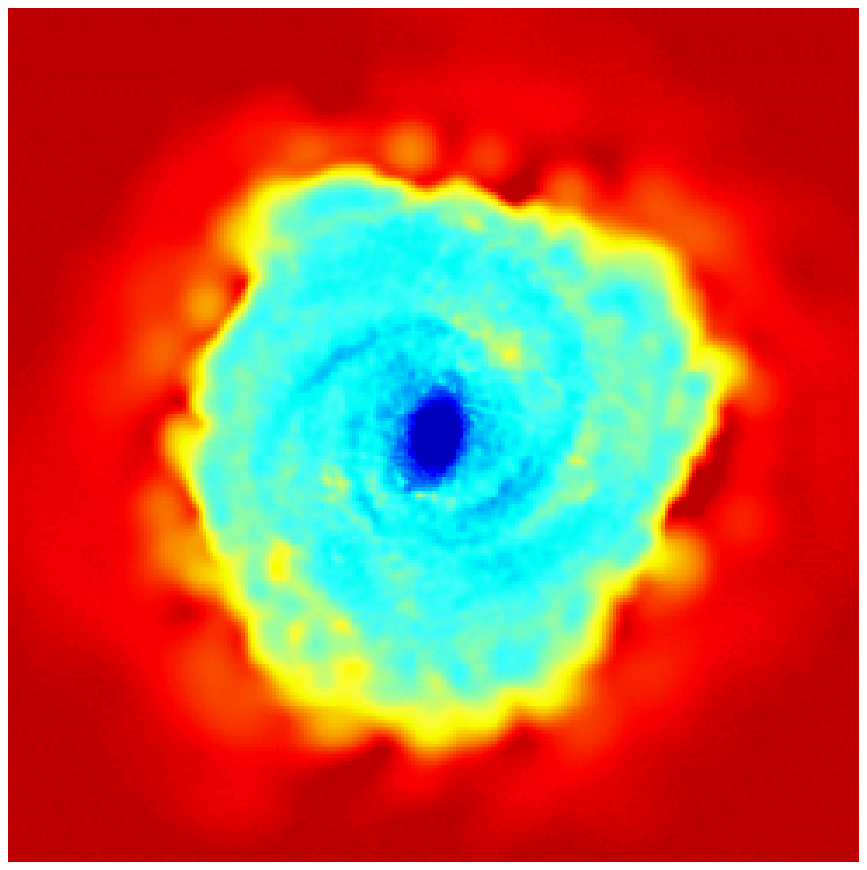}\includegraphics[%
  scale=0.614]{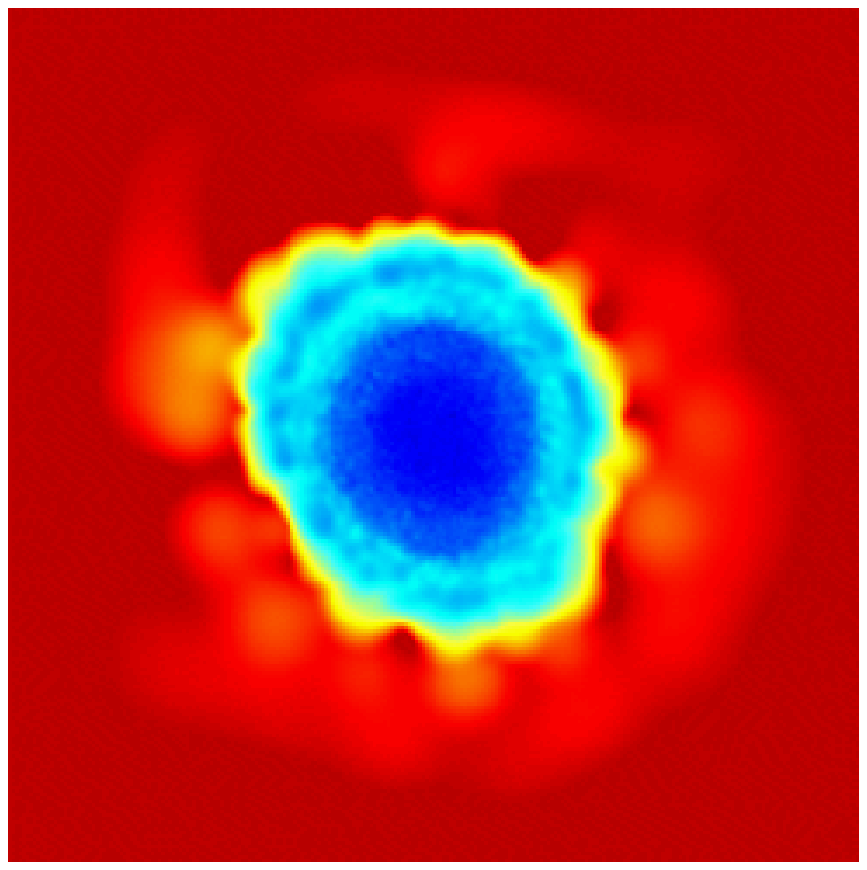} \includegraphics[%
  scale=0.32]{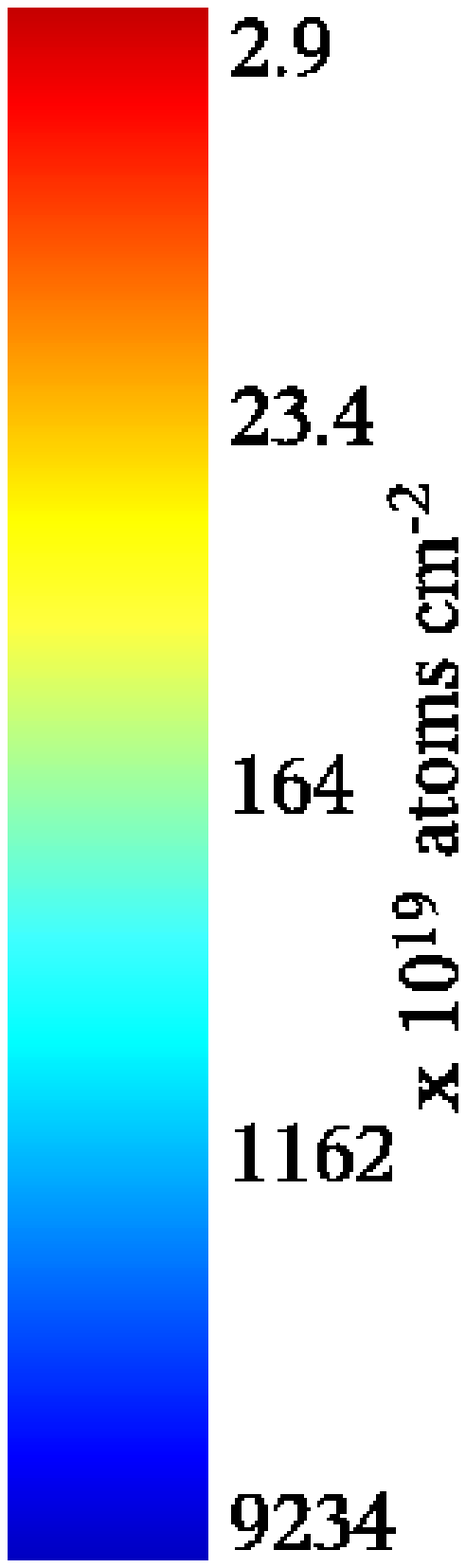}

\caption{The three panels show density maps of gas in a slice through the
centre of the Milky Way gas disk after $5$ Gyr, from left to right: 
HRLS, IRLS, LRLS. Box side length $20$
kpc for every panel - clearly the disk is larger for 
higher resolution and the bulge to disk ratio lower ($0.72, 0.80$ and $0.90$ for HRLS, IRLS and LRLS, respectively). \label{fig: picts of gasdisks}}
\end{figure*}

\begin{figure}
\includegraphics[%
  scale=0.4]{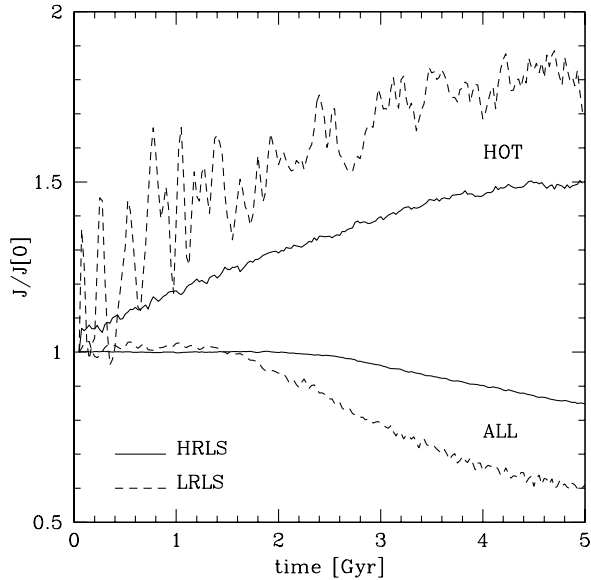}

\caption{The time evolution of the specific angular 
momentum for gas particles which were initially in a shell from 30 kpc to 40 kpc around the centre is shown. All gas particles from that shell of HRLS and  gas particles from the same
shell which remain hot during the simulation have been measured, and the equivalent for LRLS. At lower resolution the loss of angular momentum in the cold phase is larger than for high resolution.
\label{Jcold+hot}} 
\end{figure}

We find that even with more than $10^5$ gas particles convergence in the final 
angular momentum is not reached. In fact the disk in the high resolution HR runs (with $500,000$ 
gas particles) accreted $\sim30\%$ more specific angular momentum than LR and still $\sim13\%$ more
than the IR run, see Figure \ref{cap:Accretion-of-J/M}.  
Over the 5 Gyr of evolution, the simulations conserve total angular momentum
 better than one part in a hundred. The particles which are in the disk after $5$ Gyr start with an initial specific angular momentum of $\sim 500$ kpc km/s in the hot phase and lose as they cool to the disk then $20\%,$ $33\%$ and $41\%$ in the HR, IR and LR simulations, respectively. The hot component gains $30$  $(15)\%$ and the dark halo gains $10$ $(5)\%$, in the LR and HR runs (the latter in parentheses). Correspondingly the size of the central disk also varies with resolution; within the LR runs it measures $\sim 4$ kpc whereas we find $\sim 7$ kpc for the HR run, see Table \ref{cap:Results-of-the-sim} and Figure \ref{fig: picts of gasdisks}. 
This exchange of angular momentum between hot and cold gas particles is also visible in a slightly different way in Figure \ref{Jcold+hot}, where we followed the angular momentum of either hot or all gas particles in a given shell at different resolutions. From thus we see that the hot phase ($T>80,000$ K) gains whereas all the gas particles from that shell and therefore also the cold phase ($T<80,000$ K) loses  angular momentum, and that this depends
sensitively on resolution.

\begin{figure}
\includegraphics[%
  scale=0.4]{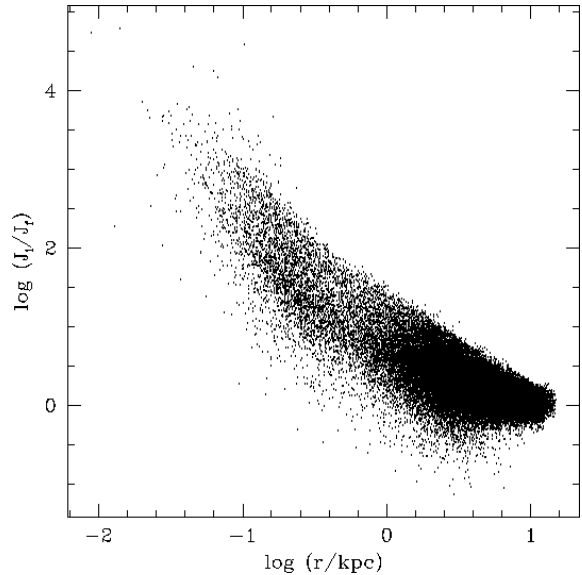}

\caption{ For the gaseous disk particles in the M33A model we plot the initial specific angular momentum 
over the final (measured at $4.8$ Gyr) versus radius. The angular 
momentum loss in the centre can be very large. \label{Jloss}}

\end{figure}

\begin{figure}
\includegraphics[%
  scale=0.4]{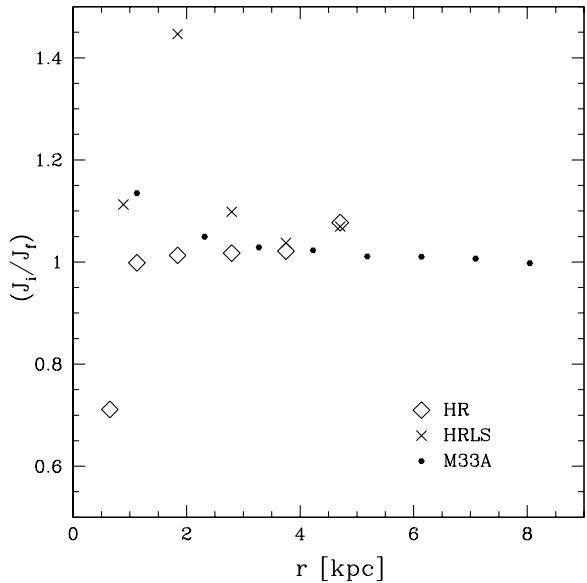}

\caption{ The ratio of the specific angular momentum of disk particles after $2.8$ Gyr ($J_i$) and after $3.8$ Gyr ($J_f$) is plotted versus radius.  The innermost point of the HR run has already lost most of its angular 
momentum before  $2.8$ Gyr during the bar instability phase at $\sim 1$ Gyr, therefore the gain here is 
insignificant.  In the HRLS run the spiral arms are redistributing angular momentum in the whole 
disk whereas for the M33A and HR run the angular momentum in the outer part of the disk is 
roughly conserved. \label{rings}}
\end{figure}

Most of the angular momentum loss occurs in the central disk. Despite the quite modest overall loss of angular momentum 
of cold particles in the HR run, the particles
which end up in a sphere of $1$ kpc radius around the centre have lost $\sim80\%$ of their 
specific angular momentum over $5$ Gyr. This loss can be even larger for particles 
which end up in the central region of the galaxy in the runs of the M33 model 
(see section \ref{M33section}). We plot the initial specific 
angular momentum of every particle over the final (at time $4.8$ Gyr) 
in Figure \ref{Jloss}. The overall loss in 
the M33A run was $14\%$, whereas for this inner $1$ kpc region $88\%$ of the specific angular momentum 
was lost. Another way to look at this is to compare the angular momentum evolution of different rings 
in the disk around the centre, which is done in Figure \ref{rings}.

Gas particles can lose  angular momentum not only in the central disk but also while they are still part of the hot halo: a substantial fraction of the particles which end up in this inner region lose angular momentum already in the hot phase.
For those particles we measured the evolution of their angular momentum while they were in the hot phase. These particles lose  13 (6)\% , in the LR and HR runs (the latter in parentheses) in the first Gyr.
The reason for this transfer might be due to the hydrodynamical drag that particles
and/or clumps of particles suffer as they move through the hot background. 
Gas particles will cool and form small clumps where there are fluctuations in density, and these
fluctuations will be larger in low resolution simulations, perhaps explaining why
the angular momentum loss is bigger than in the higher resolution simulations.

It is remarkable that we only find convergence in 
terms of the accreted cold mass but not in terms
of angular momentum and disk sizes.
In fact cold gas particles in LR runs end up with only 79\% of the angular 
momentum attained in HR runs (the difference in disk sizes is at 
the $67\%$ level, where the radial extent of the cold gas particles has been measured).

We also investigated the role that the artificial viscosity  can have in the
spurious dissipation of angular momentum. In all the runs we adopt the standard Monaghan form of the viscosity (Monaghan 1992), and in most of them
we followed Balsara (1995), who proposed to include a term to reduce excessive
dissipation of momentum in shear flows such as those present in differentially
rotating disks. 
Without the Balsara term the angular momentum of the disk for the IR run grows less
without it, see Figure \ref{cap:Accretion-of-J/M}.
The difference between the two cases is comparable to the difference
between low and intermediate resolution, implying that the Balsara
term is still important as long as the number of gas particles is of order
$10^5$. Similar conclusions for lower
resolution runs were found in previous works, for example Thacker et al. (2000).

\subsection{Torques and losses from smoothing and artificial viscosity\label{sub:Torques-at-Different}}

Okamoto et al. (2003) have pointed out, that there is significant transfer of angular momentum from the 
cold gas disk to the ambient hot halo gas due to spurious pressure gradients across the interface between
the two phases. These gradients arise because hot particles belonging to the diffuse halo but located near 
the edge of the much denser disk have their density overestimated via the SPH smoothing calculation. This problem has led some authors to explicitly decouple the cold and hot phase in the SPH smoothing (Pearce et al. 1999; Marri \& White 2003). 

\begin{figure*}
\includegraphics[%
  scale=0.4]{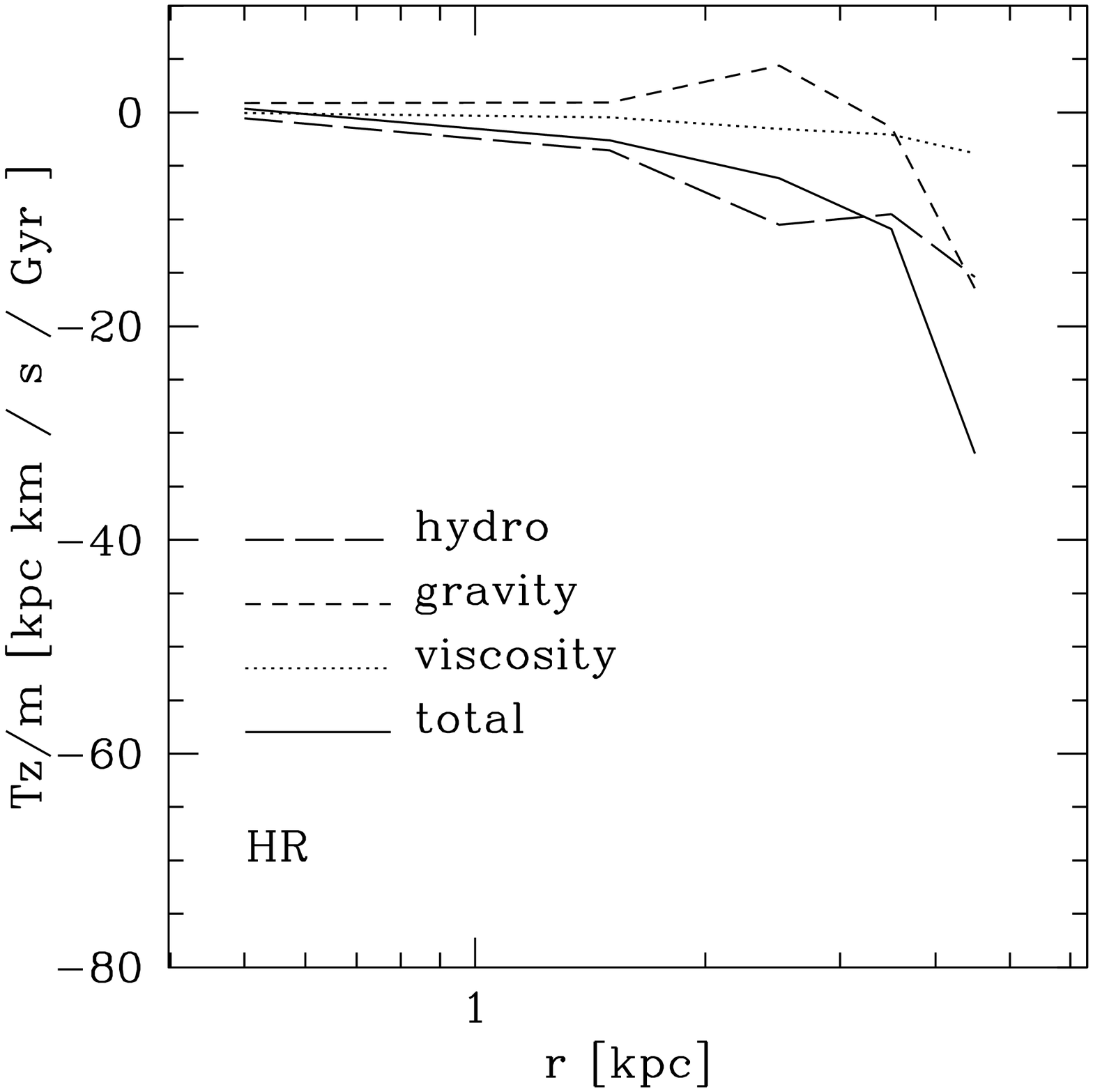}\includegraphics[%
  scale=0.4]{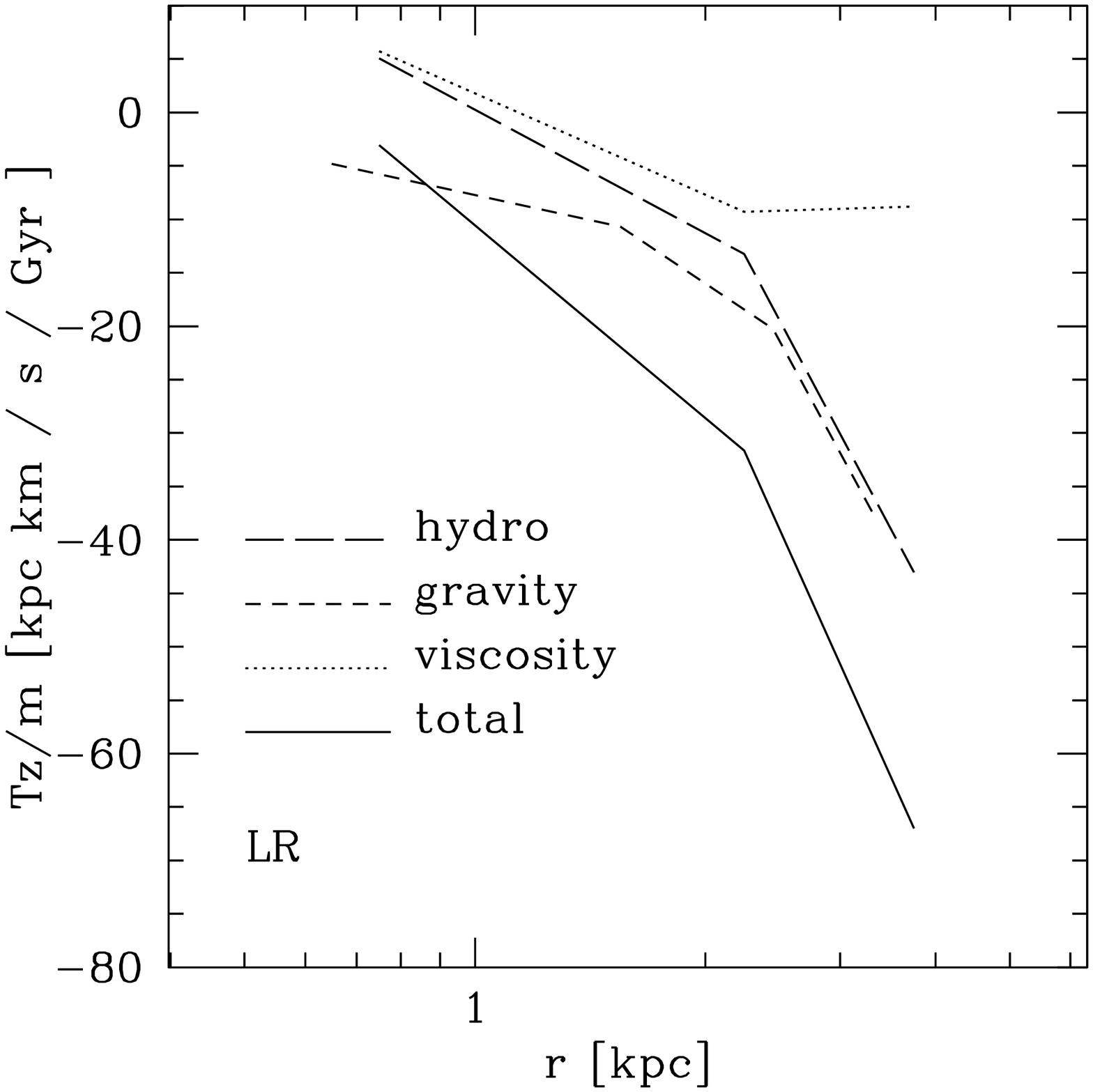}
\includegraphics[%
  scale=0.4]{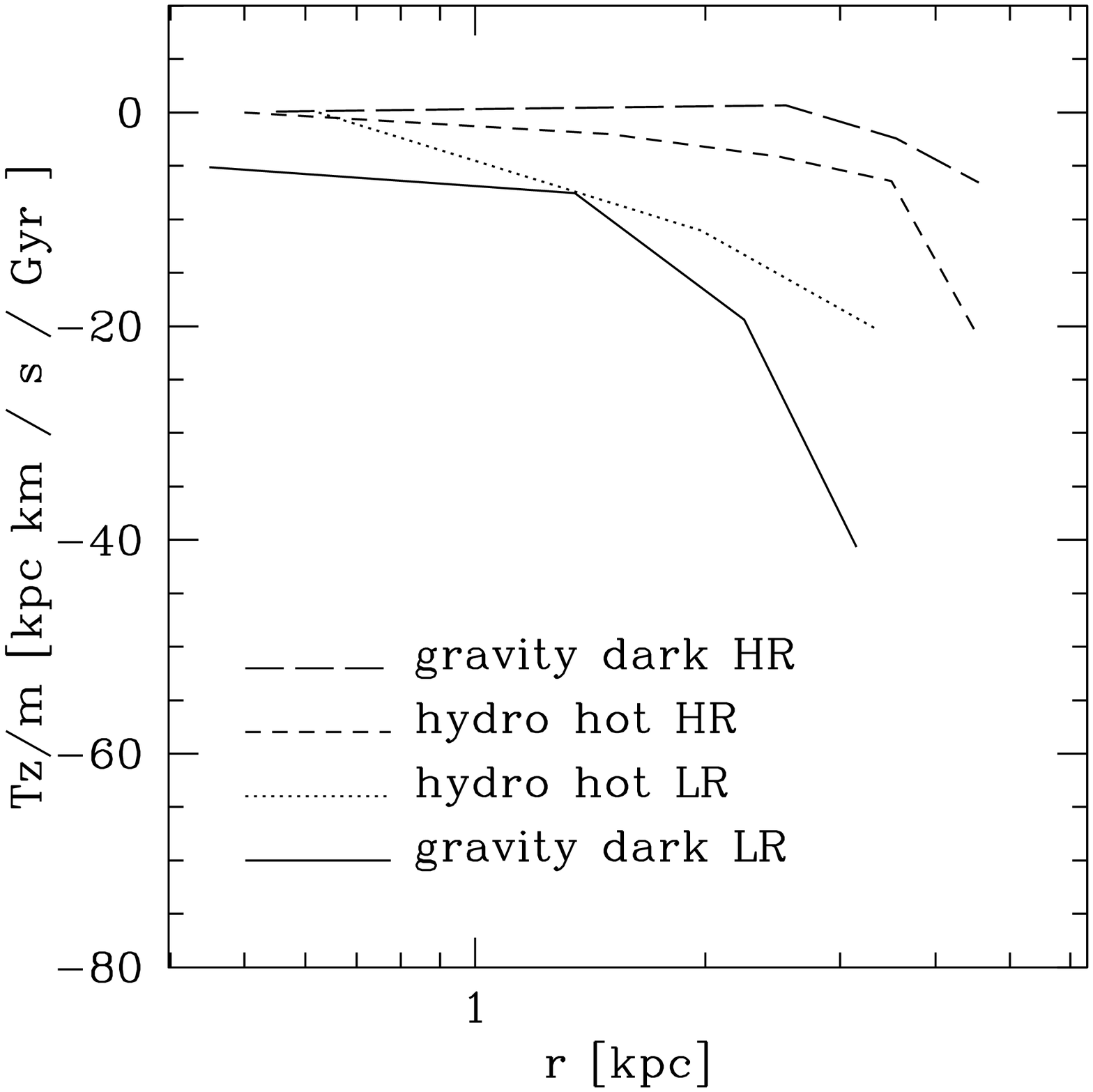}
\includegraphics[%
  scale=0.4]{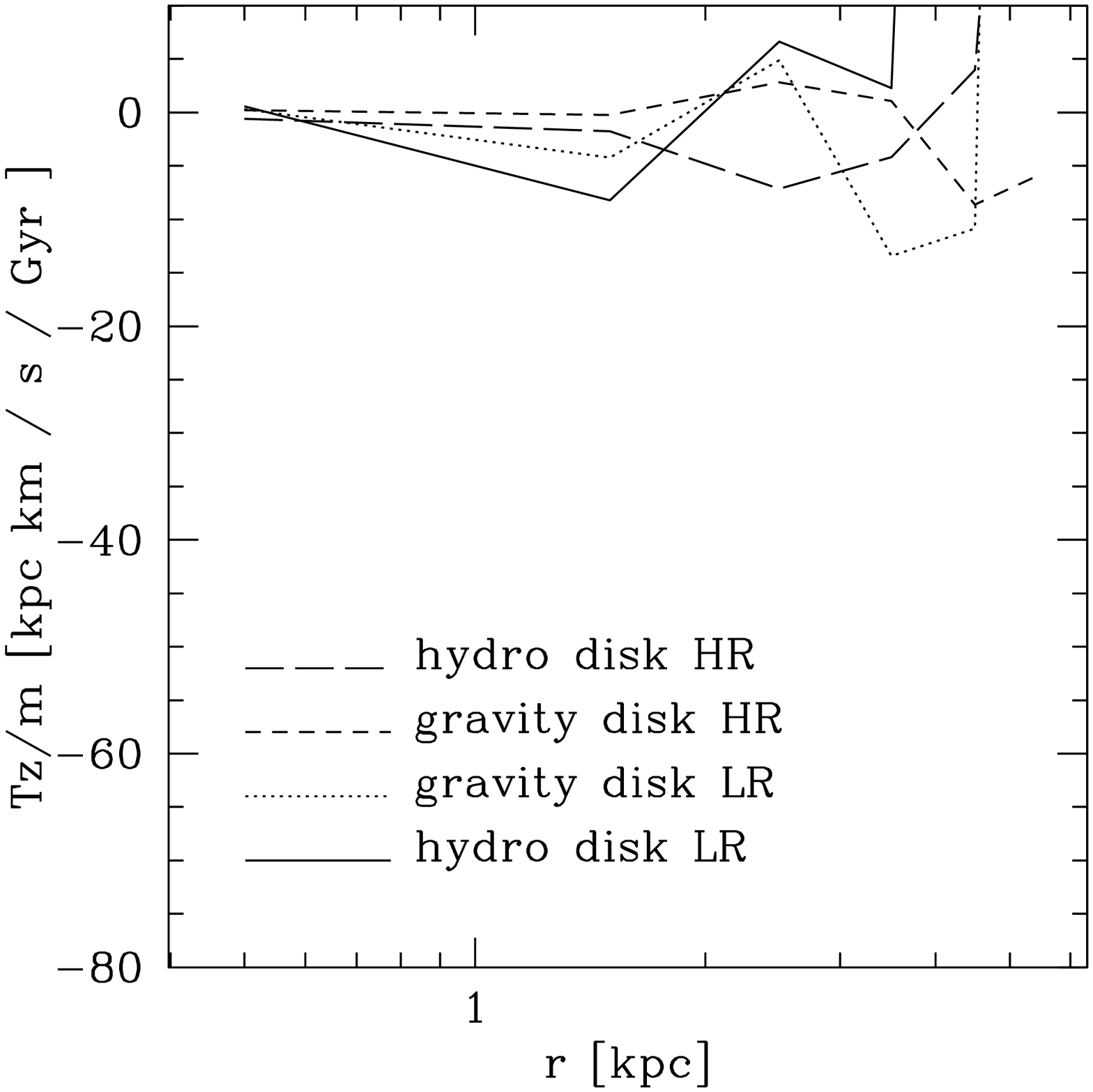}
\caption{
Azimuthally averaged specific torques (after 5 Gyr) parallel to the angular momentum of the disk are plotted as 
functions of radius in the upper panels:  hydro torques, gravitational torques, artificial viscosity torques  and total torques. In the lower left 
panel we plot the hydro torques between the hot gaseous halo and the cold gas disk. Furthermore the gravitational torque from the dark halo is shown for the LR and HR simulation, respectively.  In the lower right panel we plot the gravitational and hydro
torques from the disk particles acting on themselves for the  HR and LR simulation. In all the plots we find significantly 
lower torques for the high-res simulation. At high resolution, pressure torques are small but dominate over gravitational and viscous torques.
\label{torques}
}
\end{figure*}

We now examine the torques which may cause the transfer of angular momentum 
between the cold and the hot phase, including their resolution dependence. We measured 
separately the torques arising from, respectively, gravitational and hydrodynamical 
forces (including both pressure and artificial viscous forces) acting on disk particles after 5 Gyr
of evolution. First of all one should notice that in the low resolution runs (e.g. LR) both the gravitational 
and hydrodynamical  torques are always significantly
larger than those in the high resolution runs (HR), as shown by the two 
upper plots of Figure \ref{torques}. Gravitational and hydrodynamical torques are also comparable at a given resolution.
The total hydrodynamical torques normalised by the 
angular momenta of the cold gas disk $(T/J)$ are $-0.13$ and 
$-0.02$ Gyr$^{-1}$ for the LR and HR run, respectively. These numbers 
are much smaller than the values ($\sim -0.9$ Gyr$^{-1}$) found  in the 
paper of Okamoto et al. (2003). This  is not surprising since even the 
LR run has $\approx5$ times more gas particles in the disk 
than in the (cosmological) simulation of Okamoto et al. (2003).

We then compute separately SPH forces arising from the cold gas (i.e. from
the disk itself) and those arising from the hot gas component. For high 
resolution (HR) the torques between the hot and the cold 
phase are quite small but together with the losses in the hot 
phase and during the bar instability, see section 4, are 
enough to explain the measured angular momentum loss
(assuming the magnitude of these torques is typical for the entire 5 Gyr of evolution)
whereas at low resolution (LR) the same torques are much bigger 
and there is a significant additional contribution from gravity torques 
(upper right plot of Figure \ref{torques}). Indeed, the 
gravitational interaction between the cold phase
and the dark halo also causes angular momentum transport 
in the LR simulation; the solid lines in the lower left plot 
of Figure \ref{torques} indicates that the 
torques from halo particles on disk particles are of the 
same order as the hydro torques. The disk did not grow exactly 
at the centre of mass of the dark halo (the misalignment was of 
order of the softening length) and and the dark halo was less axisymmetric compared to the HR run. The latter asymmetry  appears to be the origin of this torque. In the lower 
right plot of  Figure \ref{torques} we examine the hydro and gravitational 
torques acting on disk particles and produced by disk particles themselves. 
Again at higher resolution the torques are smaller. This reflects also the 
more axisymmetric disk in the HR simulation. In any case the magnitude of these torques 
is always smaller than the torques from the halo (both gaseous and dark).

Finally, we also compared the torques coming from artificial viscous forces with the overall
hydrodynamical torques. We find that viscous torques are a small fraction of the latter at high
resolution, hence torques coming from Okamoto-type pressure gradients dominate. In the past, numerical
tests done at lower resolution were finding that artificial viscosity was probably causing most 
of the angular momentum loss in disks (e.g. Thacker et al. 2000). Our results are
not inconsistent, 
but rather they suggest that by increasing the resolution one progressively enters different 
regimes in which different numerical effects become dominant.

The results discussed  in this section cannot be simply extrapolated 
to the case of cosmological simulations of galaxy formation because the 
formation history of a galactic system is more complicated than in the models employed in this paper.  A present-day galaxy is the end result of an early phase of mergers plus a late phase of gas accretion. In the mergers phase  the angular momentum in merging objects is in 
an orbital rather than in a spin component. This early mergers phase is responsible for the formation of the spheroidal component of
galaxies. However, even in this phase, if the merging lumps are too concentrated, because they have lost too much angular momentum owing to 
the same numerical problems discussed here, they will form a bulge which is too concentrated. This certainly plays a role in the fact
that rotation curves of simulated galaxies are too steep towards the centre compared to real galaxies (Abadi et al. 2003; Governato
et al. 2006). The resolution effects that we described here thus are relevant to cosmological runs at all epochs in all objects in which
gas accretion occurs mostly from a diffuse hot component in nearly hydrostatic equilibrium.  This is indeed the dominant mode of gas
accretion for systems with masses above $\sim 10^{11} M_{\odot}$, while below that accretion in the form of cold flows becomes important
(Kere{\v s} et al. 2004; Dekel \& Birnboim 2006) and one might has to design different types of tests.   Therefore one has to conclude that 
cosmological runs are still plagued by numerical loss of angular momentum. This will be particularly severe in the progenitors of the final 
object, that have fewer particles.

\subsection{Disk scale-lengths and bar formation}

\begin{figure}
\includegraphics[%
  scale=0.8]{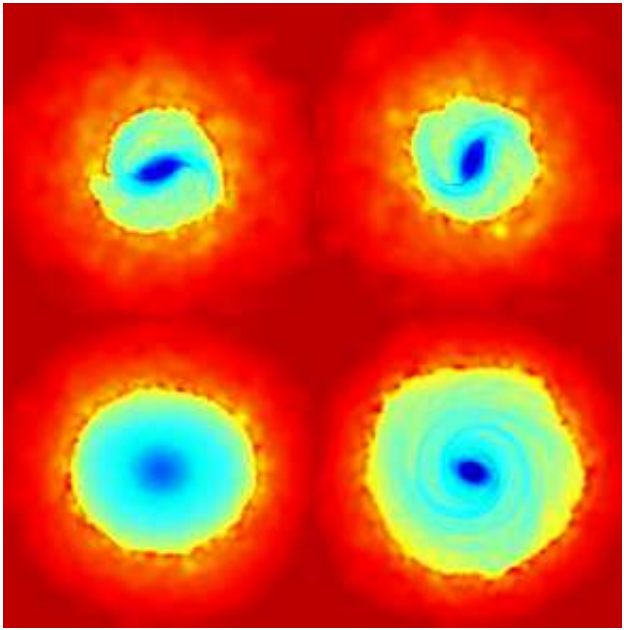}
\caption{Density maps of gas in a slice through the centre of the Milky Way gas disk which show  how bar formation can be suppressed by a large softening. On the left the HR run is shown, on the right the HRLS run. Upper panels show a snapshot after 1 Gyr, lower panel one after 5 Gyr. Boxes are 20 kpc on a side. The density is colour coded as in Figure \ref{fig: picts of gasdisks}.
\label{cap:bars}}
\end{figure}

\begin{figure}
\includegraphics[%
  scale=0.4]{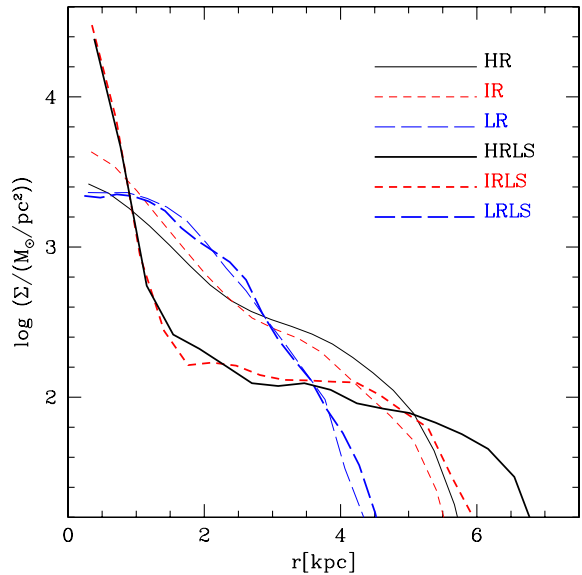}

\caption{The surface density of the gas disks after 5 Gyr. For 
small softening the Milky Way model developed a bar which increases the central 
surface density, therefore an exponential profile can not fit these models.
\label{cap:plot surfacedens}}
\end{figure}

The force resolution in the simulations is determined by the gravitational softening length and structures at scales smaller than the softening will have spurious dynamics. 
We considered the cases $\epsilon_{1}=0.5$ kpc and $\epsilon_{2}=2$ kpc - the latter typical
of that used in cosmological simulations. We found that
the choice of the softening heavily affects the morphology of the disk. A small softening allows
the formation of a bar in the central object and, afterwards, of spiral structure, 
while with the larger softening $\epsilon_{2}$ the disks remain axisymmetric and spiral structure is suppressed. While these differences are
not surprising we point out that the bar forms on scales of about 2 kpc, and therefore this is also the characteristic 
wavelength of these non-axisymmetric modes.
If the force resolution is smaller than this wavelength the modes will 
be damped, as it is the case for the large softening.
For the intermediate and high resolution case with the larger softening $\epsilon_{2}$ we notice that a bar-like 
feature first appears in the central region 
(2-3 kpc in size) about 1 Gyr after the beginning of the simulation but then
disappears shortly afterwards, see Figure  \ref{cap:bars}.   We verified that the analytic  criterion of disk stability against bar formation of  Efstathiou, Lake, \& Negroponte (1982) (see also Mayer \& Wadsley 2004) is violated in all runs that form a bar. The bar strongly affects the evolution of the radial surface density profile  (see Figure  \ref{cap:plot surfacedens}) because it triggers
the transport of angular momentum from the inner to the outer
part of the galaxy and hence an inflow of mass 
(see e.g. Mayer \& Wadsley 2004, Debattista et al. 2004 and Debattista et al. 2005).   Another mechanism to redistribute angular momentum is transport by spiral arms as seen e.g. in the M33 run, where no bar was present. Both of these mechanism can produce over-concentrated bulge-like components. As in the runs with with larger softening we find that the LRLS run does not match the results of the higher resolution simulations, it also produces
disks that are too small and have too much mass in the central regions, 
see Figure \ref{cap:plot surfacedens}.

Assuming an exponential surface density profile for the cold disk-material
Mo, Mao \& White (1997) calculated the disk scale-length to be: 
\begin{equation}\label{Mo}
R_{d}=\frac{1}{\sqrt{2}}\left(\frac{j_{d}}{m_{d}}\right)\lambda r_{200}f_{c}^{-0.5}f_{R}(\lambda,c,m_{d},j_{d}),
\end{equation}
where $m_d$, $j_d$ are the fraction of the total mass and angular momentum in the disk, 
respectively, and the functions $f_c$ and $f_R$ take into account the differences to the derivation 
for an isothermal sphere, see Mo, Mao \& White (1997), assuming conservation of angular momentum 
on all scales and that the gas starts
from the same distribution and magnitude of specific angular momentum as the 
dark matter. The information on the initial angular momentum content is
contained in the spin parameter $\lambda$. 
For our simulations
we calculated the scale-length $R_{d}=3.8$ kpc for $\lambda=0.038$ (this being
the value adopted in our MW model). From the simulations we find
$R_{d}=1.4$ kpc for the HR run,  $R_{d}=3.0$ kpc for the outer
part of HRLS, and $R_{d}=0.8$ kpc for the LR run.
 We find that the gas which settles into the disk after $5$ Gyr is cooling inside a
sphere with  maximum``cooling radius'' $r_{c}=70$ kpc in, for example,
run HR. Although one could associate $r_{200}$ with $r_{c}$, thereby matching the scale length for HR, the results of HRLS (here the central bar increases the scale-length of the outer part due to the transport of angular momentum) support the Mo, Mao \& White formula.
Clearly the measured disk scale lengths depend sensitively on the 
force and mass resolution and the comparison with analytical models remains difficult.

\section{Stellar versus gaseous disk formation}

\begin{figure}
\includegraphics[%
  scale=0.4]{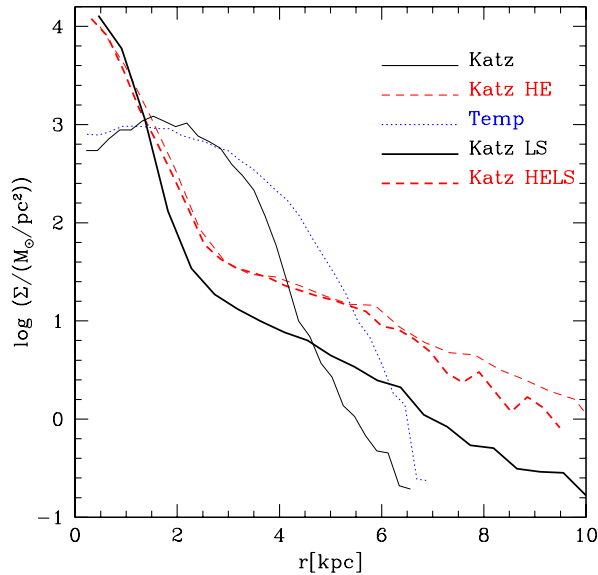}

\caption{The surface density of the stellar disk after 5 Gyr for IR resolution (see table \ref{cap:Overview-to-all-sim}, LS is referring to $0.5$ kpc softening, the other runs use $2$ kpc). 
For small softening, the Milky Way model developed a bar which increases the 
central surface density. This bar is only resolved with small softening.
 \label{cap:surfacedensstar}}
\end{figure}

\begin{figure*}
\includegraphics[%
  scale=0.4]{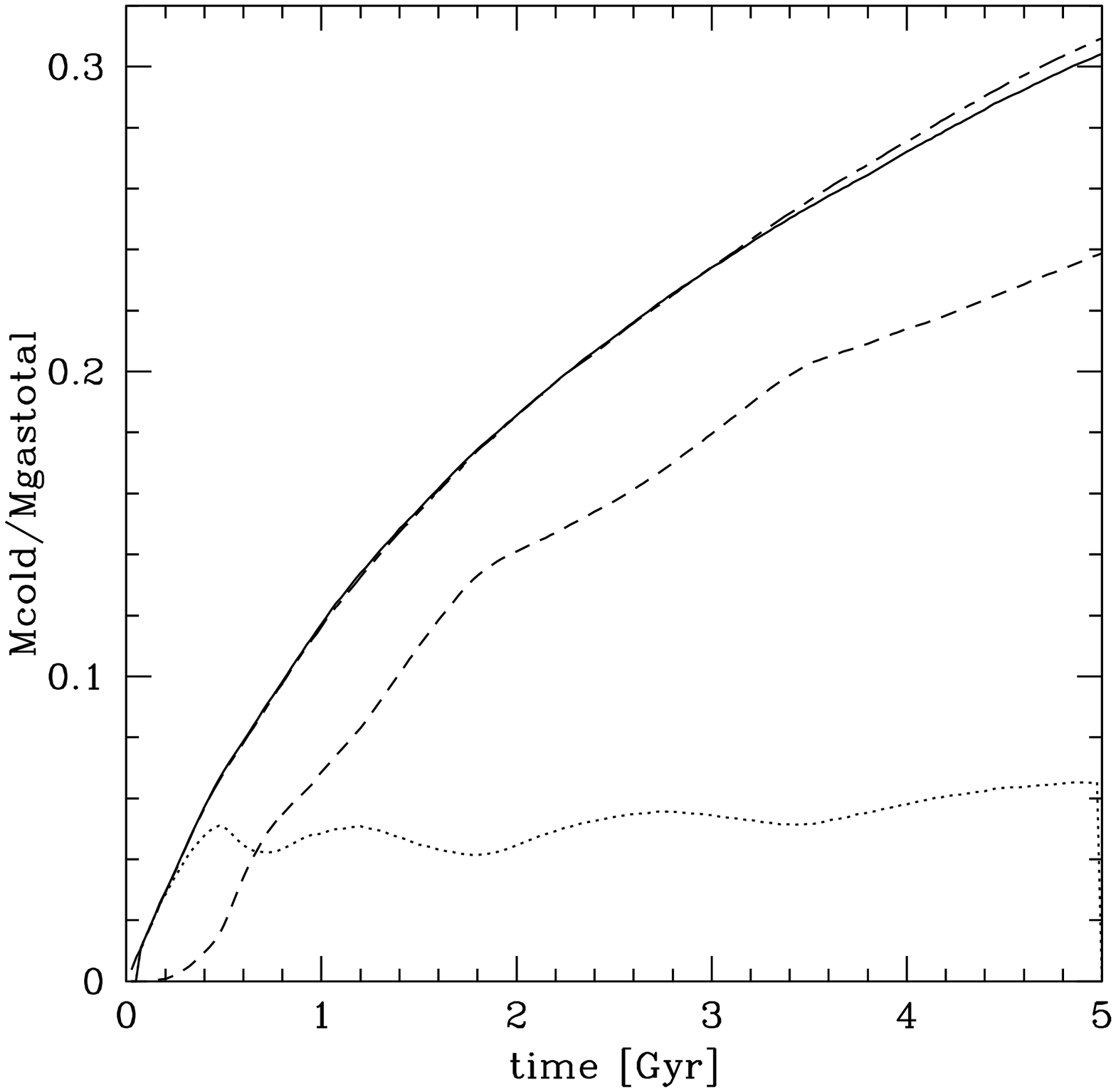}\includegraphics[%
  scale=0.4]{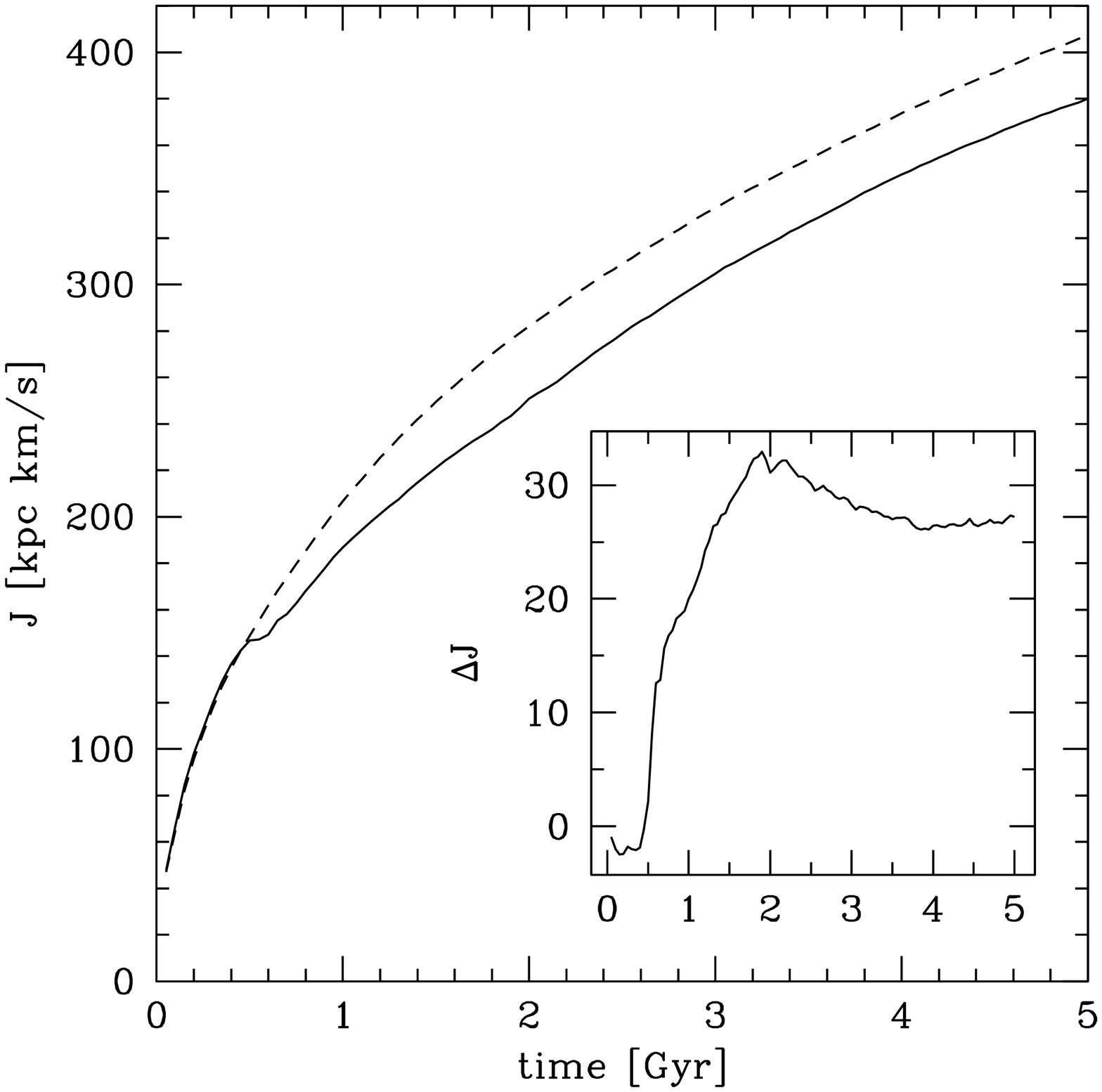}

\caption{Angular momentum and mass accretion in the simulations including
star formation and run with low softening: In the left panel we plot the 
mass accretion of the disk particles
of the star formation run with the Katz recipe (dashed stars, solid
baryons and dotted gas) and long-dashed-short-dashed 
the corresponding gas run IRLS: the accreted baryonic 
mass in the disk remained the same.  The right panel shows 
the specific angular momentum evolution of the disk particles
of the HRSFT star formation run (dashed) and the corresponding 
high-res gas run HR (solid line). In the inset the difference in 
angular momentum between these runs is plotted. The amount of angular momentum accreted by the disk in the star formation run was slightly larger compared to the purely gasdynamical run, but see text for the details.
\label{cap:angular-momentum-stars}}
\end{figure*}

In reality star formation will occur in a galactic disk as soon as it begins
to assemble. In fact in absence of the heating provided by feedback from star formation
the gas layer will rapidly become gravitationally 
unstable, dense molecular gas will form efficiently and 
will collapse into star clusters. We 
performed some additional runs with star formation. These 
runs will be used to isolate more clearly that component 
of spurious angular momentum loss which is related only 
to the SPH modelling of gas physics. We use the fact that if the gas in the disk is quickly turned into stars it will interact with the rest of the system only via gravity. Among the
star formation prescriptions that we have adopted, some are not
meant to be realistic but are just designed to test such numerical issues. Unless 
stated otherwise these tests have been run with the Milky Way model at intermediate resolution (IR).
\label{Starformation}
We began by using a star formation recipe similar to that described 
in Katz (1992); stars spawn from cold, Jeans unstable gas particles in 
regions of converging flows. The mass of gas particles decreases gradually
as they spawn more star particles. After its mass has decreased below 10\% of 
its initial value the gas particle is removed and its mass is re-allocated 
among the neighbouring gas particles. Up to six star particles are then created 
for each gas particle in the disk. Note that once a gas particle is
eligible for spawning stars, it will do so based on a probability 
distribution function.  The star formation rate is $4.5$  
$M_{\odot}yr^{-1}$ (time-averaged over $5$ Gyr) in the run adopting this recipe and the resulting disk mass 
is very close to that of the disk in the corresponding 
run without star formation (see Figure \ref{cap:angular-momentum-stars}). 
Moreover, similarly to the purely gas dynamical runs, a bar
forms provided that the force resolution is high enough. Therefore the
stability properties of the disk seem essentially independent to its
collisionless or dissipational nature, probably because the disk is massive
enough to make velocity dispersion or disk temperature factors of secondary 
importance (Mayer \& Wadsley 2004). Also the stellar surface densities are 
close to those of the gas runs, as shown in Figure \ref  {cap:surfacedensstar}.

\begin{figure*}
\includegraphics[%
  scale=0.5]{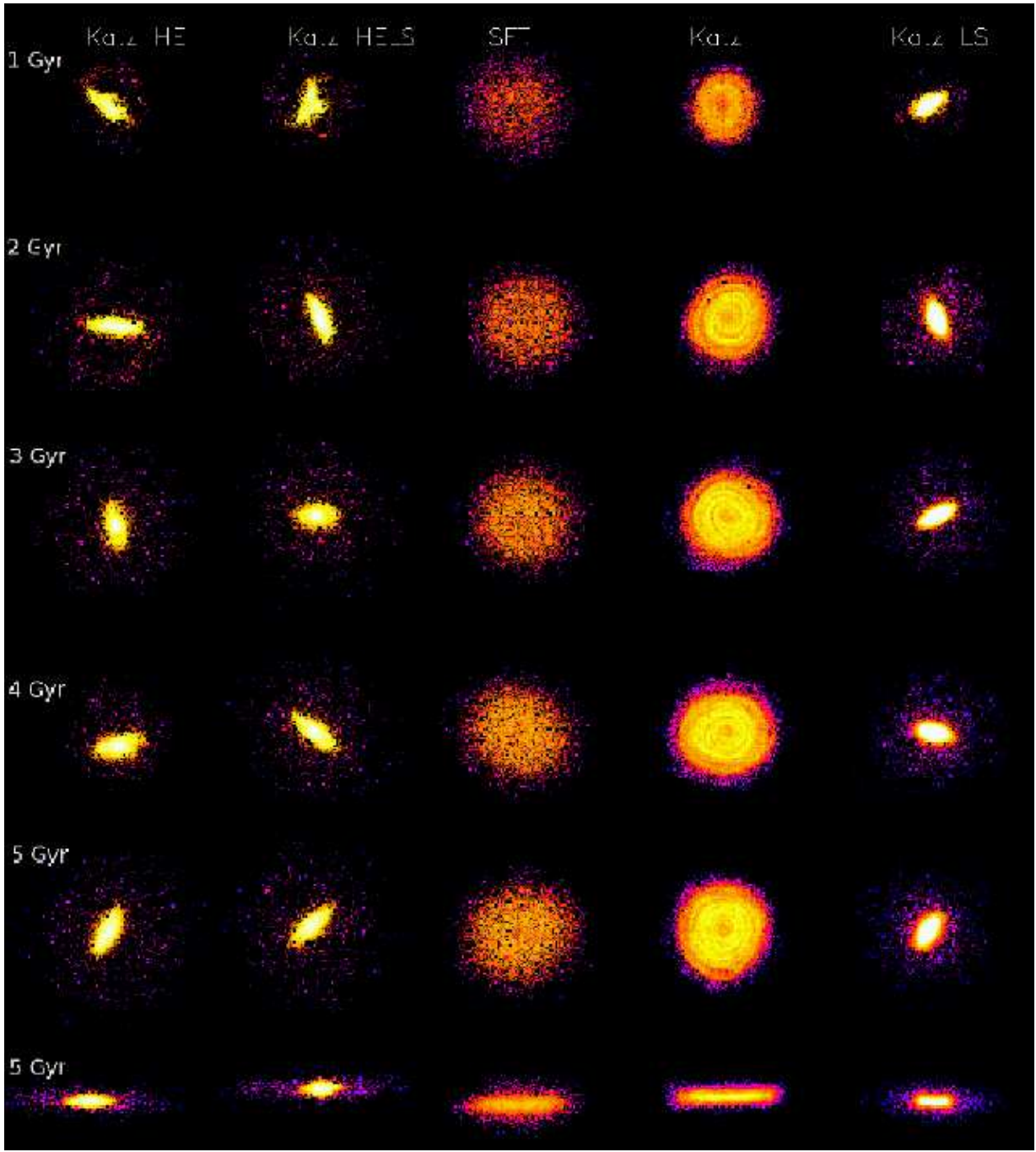}

\caption{Colour coded density maps of the distribution of stars - each 
column is a different star formation recipe and/or softening. Column 1
and 2 
show the runs with the high efficient, Katz-type star formation recipe (Katz HE, see Table 1 and 3), while column 4 and 5 use the standard Katz star formation recipe (see Table 1 and 3). In column 3  the star formation was based only on a temperature criterion (runs with extension ``SFT'' in Table 1 and 3). Boxes are $16$ kpc on a side for each individual snapshot. In all the runs with  small softening (second and last column) the bar-instability was resolved.\label{cap:starpics}
The final row shows the end state seen edge on.
}
\end{figure*}

We then explored a second star formation recipe in which whenever a gas 
particle satisfies the relevant criteria (i.e it is in a convergent flow, 
the condition of local Jeans instability is satisfied and the temperature is 
lower than 30,000 K) is immediately turned into a single star particle of the same mass. 
In this way the efficiency of star formation is much higher than in the standard scheme.
We refer to this model as Katz HE. The star formation rate is slightly higher here
($5$  $M_{\odot}yr^{-1}$) (time-averaged over $5$ Gyr), but the initial star formation rate was much higher, $11.1$ $M_{\odot}yr^{-1}$ time-averaged over the first $0.5$ Gyr, compared to  $4.0$ $M_{\odot}yr^{-1}$ for the first model during the same time. The gaseous disk soon becomes
mostly stellar ($M_{gasdisk}/M_{disk}\approx0.15$ after 5 Gyr). The massive 
and dense stellar disk that forms so early produces a particularly strong and
long bar that this time is not suppressed even with a softening of $2$ kpc
(see Fig. \ref{cap:starpics}). 
Other than this, most of the properties of this simulation are similar to
those of others, including the total accreted disk mass.

To separate the gravitational and hydrodynamical aspects of disk formation as cleanly as possible we used a third 
star formation recipe which is based only on a temperature criterion (this run is indicated as SFT and was
carried out at HR resolution). As soon as a 
gas particle becomes colder than $30,000$ K it 
is immediately turned into a star particle. The resulting star formation
rate is much higher than star formation rates in disk galaxies, and is essentially equal to the mass
deposition rate in the cold phase, ($\sim 6.3$  $M_{\odot}yr^{-1}$ is the time-averaged value
over $5$ Gyr). 
In this run there is never a cold gaseous disk, instead a purely stellar disk
forms as soon as gas particles cool down.
This means that, by construction, there are no torques between a hot and a cold phase. The disk has
a bar when the softening is small ($0.5$ kpc). The disk of the high resolution run HRSFT accumulates $\sim7\%$ more angular momentum after $5$ Gyr 
compared to the corresponding run with no star formation HR (see Figure \ref{cap:angular-momentum-stars}). 
However, the two runs differ significantly because the run with star formation
never forms a bar (this recipe produces a relatively thick disk which has a higher Q parameter)
while the HR run has a transient bar instability between $0.5$ and $2$ Gyr (both runs
have large softening). Interestingly, the difference in the final angular momentum content of the disks seems
to arise during the period in which run HR forms a transient bar (as shown by the inset plot in 
Figure \ref{cap:angular-momentum-stars}) transferring  angular momentum to the halo. Indeed if we compare the angular momentum evolution between the two runs
after $\approx2$ Gyr, when also the HR
 run has formed a stable disk, they appear virtually identical.
We argue that this shows that at high resolution the artificial hydro torques between the hot and cold phase are 
small, and hence any eventual loss of the disk is dominated by gravitational torques (torques from artificial viscosity
are even less important as we have shown previously). These results support the analysis carried out in section 3.2.

\section{Dependence on the Initial Conditions}

\subsection{Influence of the temperature floor on the gravitational stability\label{sub:tempfloor}}

We adopted a standard cooling function for a gas of primordial composition
(hydrogen and helium in atomic form) including radiative cooling which
is very efficient above temperatures $10,000$ K (molecular cooling is
not included). The central gas quickly cools down to either this limit
or whatever temperature floor we impose.
In all the models without a temperature floor in the cooling
function the disk becomes gravitationally unstable after a couple of Gyr
and widespread fragmentation into dense clumps occurs. A way to quantify
the stability of a gas disk is the Toomre parameter (Toomre, 1964).
\begin{equation}\label{Q}
Q(r)=\frac{v_{s}\kappa}{\pi G\Sigma},
\end{equation}
where $\kappa^{2}(R)=R\frac{d\Omega^{2}}{dR}+4\Omega^{2}$ is the
local epicyclic frequency, $v_{s}$ is the local sound-speed and $\Sigma$
is the surface density of the gas. For a razor-thin gaseous disk, stability
against all axisymmetric perturbations requires $Q>1.$ The disks fragment
because $Q$ drops to less than unity almost everywhere in the disk shortly
after formation.

\begin{figure}
\includegraphics[%
  scale=0.4]{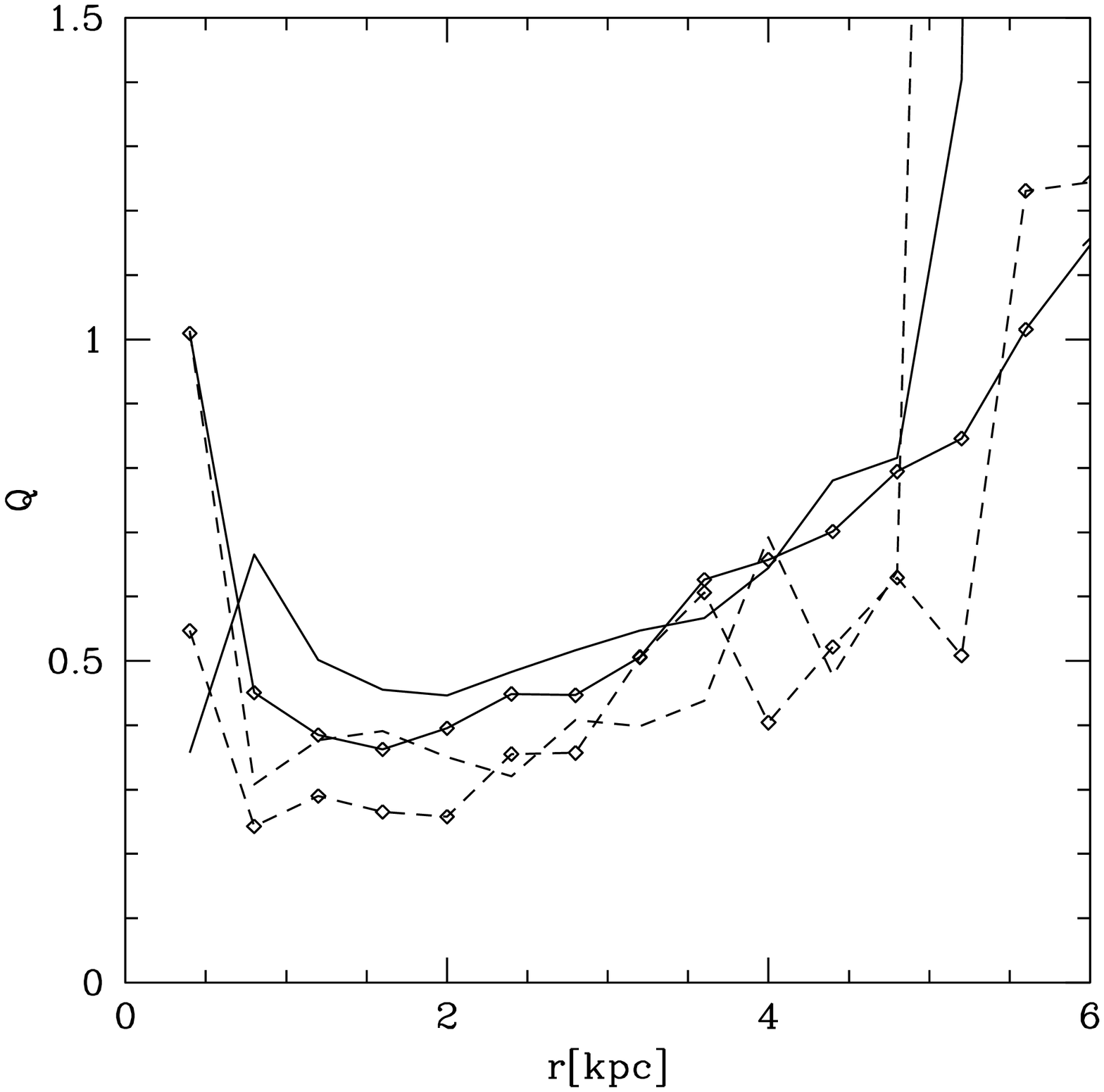}
\caption{Plot of Toomre parameter Q for the high-res model with (HR, solid lines) and without (dashed lines) a temperature floor in the cooling function. 
The lines with open symbols are after $6$ Gyr whereas the other lines are after $2$ Gyr. 
As expected, Q increases if ones uses a temperature floor.\label{toomre}}
\end{figure}

To prevent these instabilities  we set a minimum temperature that the gas can reach by cooling. In reality heating from stellar feedback (stellar winds and supernovae
explosions) will provide a way to maintain the effective temperature of the 
disk above $10,000$ K (between $15,000$ and $30,000$ K depending on the run).
Evidence of this is the existence of a multi-phase structure
in the interstellar medium, which includes a cold, a warm and a hot phase
(McKee \& Ostriker 1977), and the fact that even the coldest phase, i.e. 
molecular gas, has significant turbulent motions which creates an effective pressure
that is higher than the thermal pressure.
Because $v_{s}\propto\sqrt{T}$, the Toomre parameter increases when we enforce such a
temperature floor, see Figure \ref{toomre}. As a result, in the run HR a fairly axisymmetric disk galaxy formed 
which did not fragment. Romeo (1994) has emphasised the
role of a finite disk thickness in the context of stability; for a thick gas disk, the effective critical value of Q is lowered down from $1$ to $\approx 0.6,$ ; the
HR run and the HR run without a temperature floor, have typical values of $Q$ that are below 1 but, 
respectively,  above and below $0.6$, thus explaining why one fragments and the other does not.
However, also the softening parameter affects stability (Romeo 1994) 
unless the disk scale-height is significantly larger than the softening. 
When the softening is important for the stability (which applies to our simulations) a cold stellar disk is stable as long as 
\begin{equation}\label{romeo}
s>s_{crit}=\frac{1}{e}\frac{2\pi G \Sigma}{\kappa^{2}}, 
\end{equation}
where $s$ is the plummer softening (which is of the order twice the corresponding 
spline softening used in GASOLINE); this result was also generalised for a gaseous 
disk. Measuring $s_{crit}$ in the simulations showed that only for LS 
runs ($\epsilon=0.5$ kpc) the softening is clearly smaller than this critical value. 
Although  Romeo's results were derived for only gravitational 
forces, this suggests nevertheless that the stability level of our cold 
gas disk is not simply controlled  by the physical properties which determine $Q(r)$, 
but is also heavily influenced 
by the choice of the softening parameter. Fragmentation and instabilities are 
suppressed by the softening, which acts as an artificial pressure at small 
scales (Bate \& Burkert 1997).
This conclusion supports the strong dependence of the bar 
instability from the choice of the softening.

\subsection{Generating initial conditions using a merger} \label{merger_section}

\begin{figure}
\includegraphics[%
  scale=0.4]{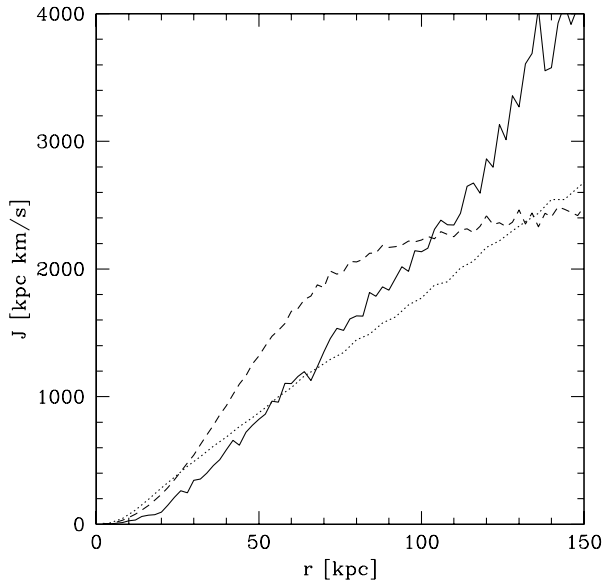}
\caption{\label{cap:surfdensJdiff} 
The specific angular momentum profile after the merger: dark matter (solid lines) and gas (dashed lines).
The dotted line show the specific angular momentum profile of the gas from the spherical MW model used previously.\label{merger}}
\end{figure}

\begin{figure}
\includegraphics[%
  scale=0.9]{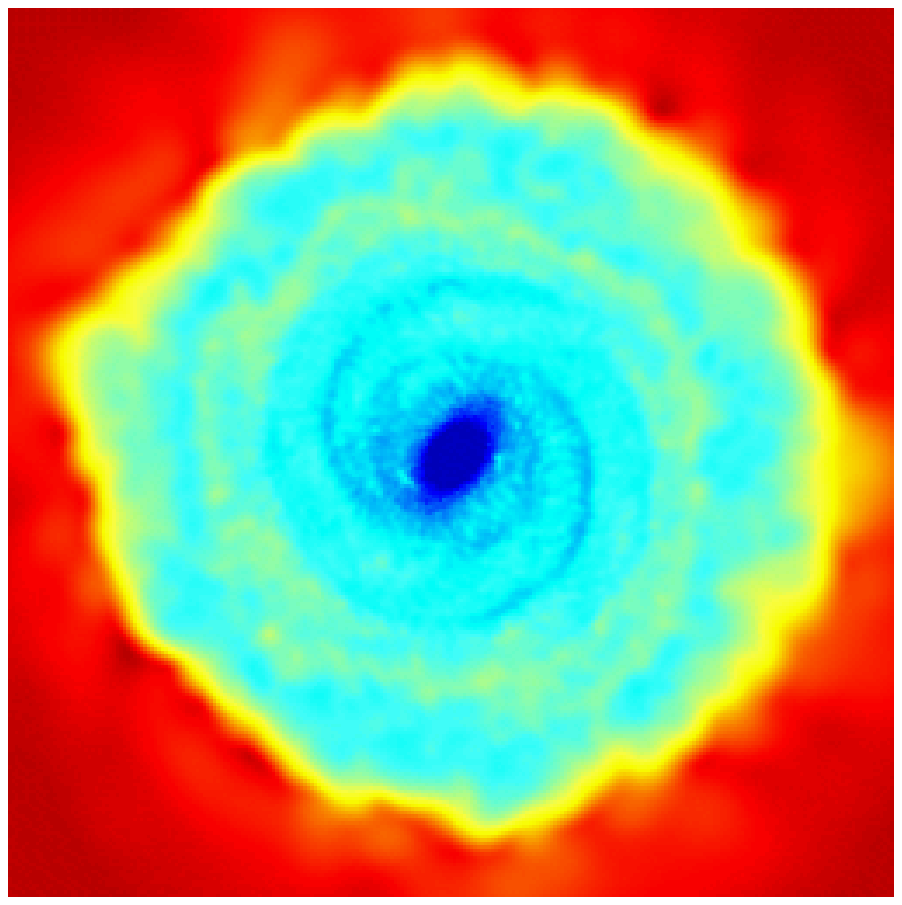}
\includegraphics[%
  scale=0.9]{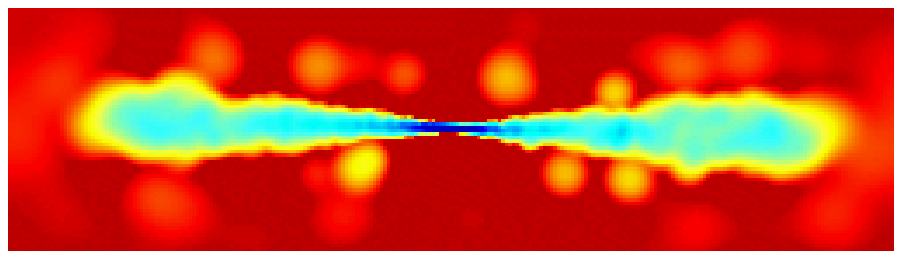}

\caption{Density maps of gas resulting from the triaxial
initial conditions in a slice through the 
centre of the gas disk after $5$ Gyr, face-on and 
head-on. The main box is $20$ kpc on a side. The density is colour coded as in 
Figure \ref{fig: picts of gasdisks}. The merger model produced a 
disk remarkable similar to that in the standard spherical halo models.\label{mergerpics}}
\end{figure}

In the $\Lambda$CDM model a galaxy like the Milky Way forms after 
a series of mergers. Typically one last major merger between two or more massive progenitors
builds up most of the mass of the final system, including most 
of the gas that will later collapse and form a rotating disk (e.g. Governato et al. 2004; Sommer-Larsen et al. 2003). 
To bring a greater degree of realism to our initial conditions we construct
an equilibrium triaxial halo in which the gas acquires its angular momentum
through an equal mass merger (see Moore et al. 2004). We start with two
spherical, non-rotating NFW haloes each with half of the mass of the MW model
considered previously. 
The two halos are placed at a separation of twice 
their virial radius. One of them is given a transverse velocity 
of $35$ km/s, and then the haloes are allowed to merge. This net velocity
(determined using a trial-and-error procedure)  
results in a final gas distribution that has a similar  spin parameter
as used previously.
The triaxial halo resulting after the merger, consisting of  
$9\times10^{5}$ dark matter and $10^{5}$ gas particles, was 
first evolved for $20$ Gyr with an adiabatic equation of state for 
the gas (softening parameter of $0.5$ kpc). The final spin parameter 
inside the virial radius was $\lambda=0.044$ and the resulting 
specific angular momentum profile had a slope that depends 
on radius, going from $j\propto r^{1.0}$ 
to $j\propto r^{0.5}$ (see Figure \ref{merger}). The inner part 
of the dark halo was significantly oblate, $c/a\approx0.57$, and the angular momentum was aligned along the short axis.
This system was then evolved for $5$ Gyr with radiative
cooling. A disk with a bar formed, with the final cold gas mass 
being comparable to the other Milky Way 
models. However, owing to the shallower inner slope of the angular 
momentum profile (and the slightly higher spin parameter) the final disk angular momentum
was $\approx30\%$ higher
than in the corresponding standard run at the same resolution (the 
IRLS run). The disk radius reached $8.4$ kpc after $5$ Gyr.
The structure of the disk in this merger simulation is very 
similar to that of the disks in the standard simulations 
(see Figure \ref{mergerpics}). 
The model violated also the  criterion of disk stability against bar formation of  Efstathiou, Lake, \& Negroponte,  but the bar instability was delayed compared 
to the IRLS run; the bar appeared after about $2$ Gyr instead 
of after just $1$ Gyr, probably because the higher angular momentum
near the centre of the system slowed down the build up of a dense, unstable 
inner disk. After $2$ Gyr the cooling of baryons made the halo rounder
(see Kazantzidis et al. 2004); the bar formed at a time when $c/a$ 
of the dark halo was already $\approx0.8$. 

\subsection{Different Initial Angular Momentum Profiles}

\label{M33section} 
\begin{figure*}
\includegraphics[%
  scale=0.7]{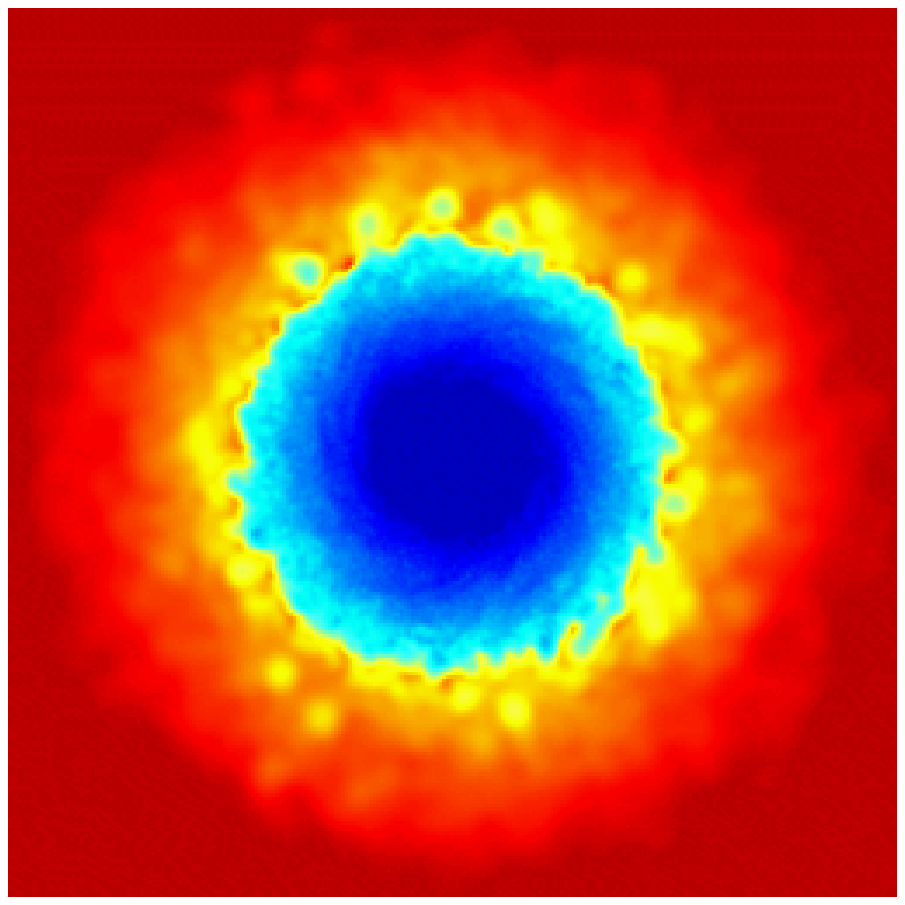}\includegraphics[%
  scale=0.7]{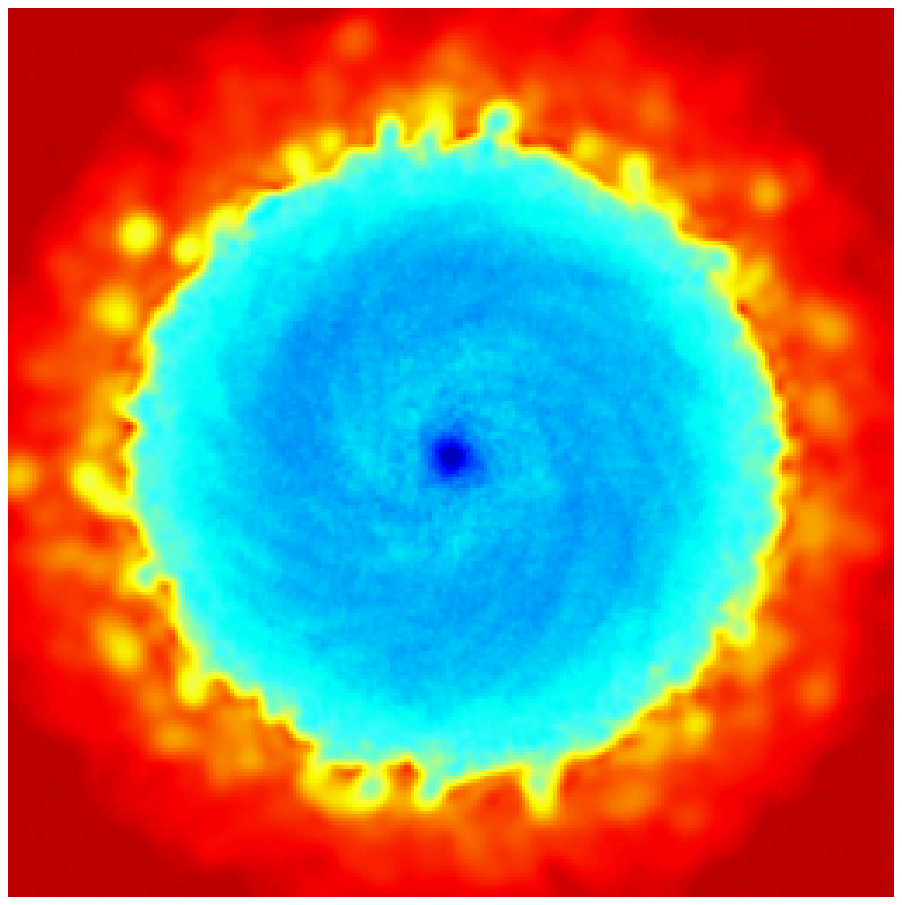}\includegraphics[%
  scale=0.3797]{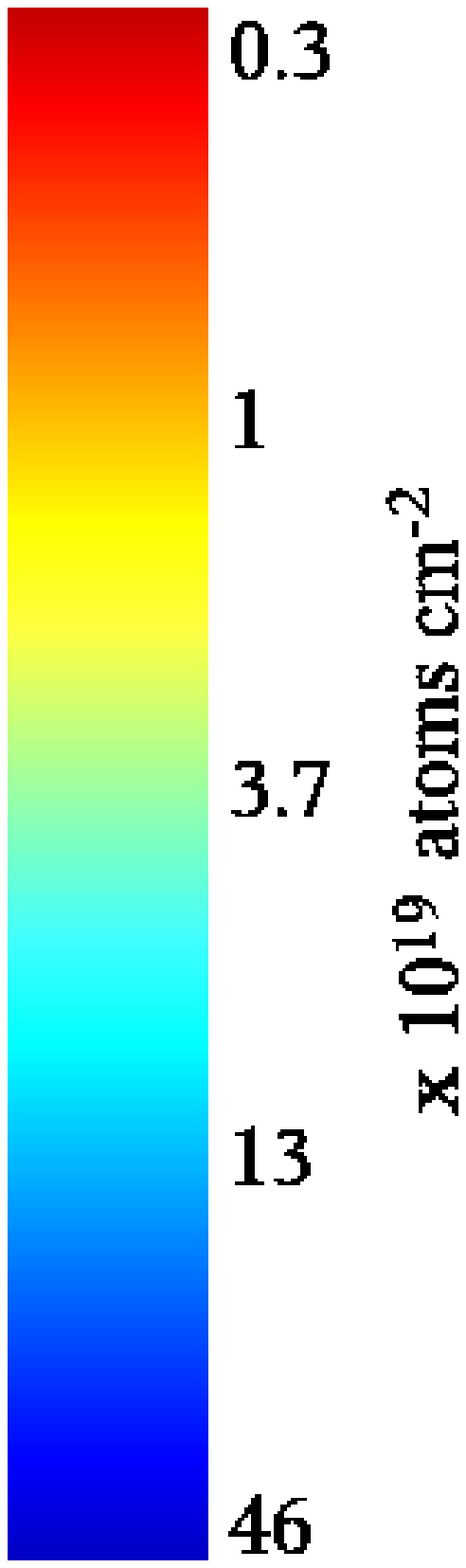}

\caption{The panels show density maps of gas in a slice through the centre of the M33 gas disk after $4$ Gyr,  for M33A (left) and M33B (right). Boxes are $50$ kpc on a side in both panels. The nucleus and the spiral arms are visible in both  models. \label{M33pics}}
\end{figure*}

As we have seen in the previous section, there are natural ways in which
the initial gaseous angular momentum profile might differ from the simple
power law as suggested by Bullock et al. (2001b).
Therefore we decided to explore further the effect of changing the 
initial angular momentum profile.  To this aim we ran an additional simulation
with the M33 model.
Here we compare runs using two different initial conditions, one with $j\propto r^{1.0}$ and one 
with $j\propto r^{0.5}$, named respectively, M33A and M33B. 
Let us first describe the results of the various runs performed at different
resolutions with model model M33A.
After 1 Gyr of evolution the surface density profiles of the standard 
resolution run and of the higher resolution, refined run with particle splitting 
(see Kaufmann et al. 2006 for the latter)
look similar, except that the 
steepening of the slope towards the centre is slightly more pronounced in 
the run with splitting, possibly
as a result of a more intense mass inflow driven by spiral arms. 
After four Gyr the disk of M33A  attains a near exponential 
surface density profile over a large fraction of its extent, except within a few hundred
parsecs from the centre, where a dense nucleus produces a central 
spike in the profile.
Model M33B also shows a spike in the centre, and the 
density in the outer region does not decrease monotonically but 
has a bump (see Figure \ref{cap:surfdensJdiff}). 
The density in the central part is one order of a magnitude smaller and 
the size of the cold gas disk is larger; its radius after $4.75$ Gyr 
is $12.7$ kpc for M33A and $19$ kpc for M33B, see Figure \ref{M33pics}. These differences are expected
because there was initially more angular momentum in the inner part of 
the halo of model M33B, and this produced a more extended disk.
The material in the central spike is 
accreted gradually over time and comes from a cylindrical region which extends above and below the disk.
After $4.5$ Gyr it has a mass of $\sim10^{8}M_{\odot}$, $\sim1\%$ of the total disk
mass for model M33A. The inner nuclear region is poorly resolved with our standard force resolution, therefore we repeated run M33A
with a softening of $50$ pc. The nucleus formed in a similar way, and its radius shrank slightly, becoming again comparable
to the force resolution. This suggests the inflow will continue to even smaller scales (unless stopped by an 
opposing pressure force produced
by some  heating source) if we had higher resolution. Artificial loss of angular momentum driven by artificial viscosity is another aspect to consider. We tested its role by decreasing the coefficients of viscosity. We run again using $\alpha=0.5$ and $\beta=0$ (strong
shocks are not expected in the disk) instead of the standard values $\alpha =1$ and $\beta = 2$, and found that neither the nucleus nor the 
appearance and size of the disk is affected by lowering the artificial viscosity. 
In summary, in no simulation using the M33 model were we able to produce a pure exponential profile without a nuclear density enhancement.

 One can ask how the structural properties of the galaxy in the M33 model compare with galaxies with comparable rotational velocities ($V_{max} \sim 130$ km/s) such as the real M33 galaxy itself. The simulated galaxy has a disk extending out to 17 kpc, which is very similar to what found by Corbelli et al. (2003). Following Regan \& Vogel (1994) and Corbelli et al. (2003), we fitted the surface density profile of the galaxy with
a combination of an exponential profile and a de Vaucouleur profile. The scale lengths of the fit turn out to be, respectively, $0.5$ kpc
for the de Vaucouleur, which dominates in the inner region where the nucleus is, and $3$ kpc for the exponential. The first scale length
is extremely close to that of M33 as determined by Corbelli et al. (2003), who found $0.4$ kpc, and the mass of the nucleus within the
scale length is also comparable, $2.3 \times 10^8 M_{\odot}$ compared to $3 \times 10^8 M_{\odot}$ in the real M33. However, the disk has clearly a shallower profile in our simulated galaxy, as shown by the fact that the exponential scale length determined by Corbelli et al. (2003) is 
$< 1.5$ kpc rather than $3$ kpc. This is also reflected in the baryonic surface density, which is almost a factor of 2 lower in the simulated galaxy
compared to M33. This difference might be due to some fundamental reason, such as a different angular momentum content, or rather reflect
the limitations of our modelling. One difficulty in the comparison is indeed the lack of star formation in the disk (we have run the M33 model
with star formation but stars formed only in the central region, very likely because none of the criteria for star formation that we used
is appropriate for a low mass, low density disk such as this one). This might affect 
the evolution of the stellar profile by affecting the stability of the disk. We see a persisting spiral pattern which is responsible for transferring angular momentum outwards, a mechanism that will tend to make the disk profile shallower, similarly to what we have seen in the case of
galaxy models developing a bar. If star formation occurs in the disk it might get hotter and the stars cannot dissipate their
random motions, possibly stifling the spiral pattern and the associated transfer of angular momentum. This might preserve a steeper profile.

The M33 model represents a more ideal case than the MW models to address the issue of disk profiles because there are weaker 
morphological changes, in particular there is no bar formation, compared to the MW models.
Van den Bosch (2001) assumed detailed angular momentum conservation to follow the collapse of spinning gas and 
the resulting formation of a disk inside
an NFW halo with a semi-analytical code. His models did not produce disks with exponential profiles, contrary 
to the assumption of the Mo, Mao \& White (1998) model. Instead power law profiles with steeper slopes near the centre were the typical outcome; those steep inner slopes agree with our simulations. However, the reason behind the steeper central slope is not necessarily the same in that work and
in our work, and so the agreement is likely coincidental.

While the lower angular momentum material collapsing rapidly and efficiently near the centre of a cuspy NFW halo was the only cause
in Van den Bosch (2001), here it seems that angular momentum transport driven by spiral instabilities is associated with the appearance
of the nucleus and the associated non-exponential profile
(effects of self-gravity are of course not included in the semi-analytical models). Does this mean that an exponential profile 
would result in our simulations if the disk was perfectly axisymmetric and smooth? We defer this important question to a future paper which will 
focus on the origin of exponential profiles. For the moment we note that the angular momentum distribution in our simulations is not fixed in 
time as in Van den Bosch (2001) but changes continuously due to a combination of numerical and physical processes depending on the force
and mass resolution. This suggests that it would not be surprising 
if the resulting profiles would be different from those calculated analytically  even in absence of spiral instabilities.

\begin{figure}
\includegraphics[scale=0.4]{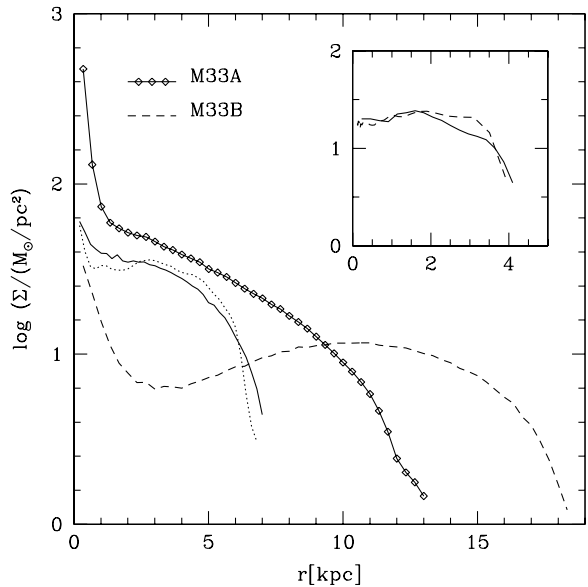}

\caption{\label{cap:surfdensJdiff}The logarithmic surface density
of the M33 gas disk after four Gyr for different models: M33A (solid with open symbols), 
M33B (dashed line). The solid line shows M33A after one Gyr and the dotted line shows the refined model of M33A at the same time.
Inset: logarithmic surface density of the standard and refined M33A (solid and dashed, 
respectively) run after $0.25$ Gyr. All the M33 models showed at later stages a nucleus in the centre.}
\end{figure}

\section{Summary and Discussion}

We simulated the formation of galactic disks in cuspy dark matter
halos to explore numerical effects and physical processes that take place during the late
stages of disk formation. We studied the effects of initial angular momentum
distribution, force and mass resolution, halo triaxiality, star formation
and artificial viscosity.
Our results show that the large angular momentum losses 
reported in some early works, including cosmological simulations,
were primarily due to insufficient mass resolution.
While convergence in disk mass is reached with $\sim 10^4$ particles,
disk sizes and angular momentum of the disk do not converge  even 
at the highest resolution used in this work ($\sim 10^6$ particles)
We studied the different numerical and physical mechanisms by which
disk particles can lose angular momentum. In summary we found the following results:

\begin{itemize}

\item Gas particles which will end up in the disk can lose angular momentum while falling  
to the disk in the hot phase. This effect depends on resolution: the low-res run loses 
twice as much as the hi-res  via this mechanism. 

\item Cold particles belonging to the disk transfer angular momentum to the hot 
gaseous halo, via both hydrodynamical and gravitational torques. The magnitude 
of these torques is comparable but depends strongly on resolution (torques are reduced by 
about a factor of 4 for an increase in resolution of more than an order of magnitude).

\item  At high resolution ($N > {10}^5$) the loss of angular momentum due to 
spurious hydrodynamical torques between the cold gas disk and the hot gas halo 
seems to play a secondary role. In fact hi-res runs in which the hydrodynamical 
interaction between the cold and hot phase is absent (where the cold gas 
particles are readily turned into collisionless stellar particles)  show 
an angular momentum evolution similar to that of the corresponding runs 
without star formation. This suggests that at high resolution the angular 
momentum transport from the disk is mostly gravitational in origin and most 
of the hydrodynamical angular momentum transport is occurring before the 
particles reach the disk (the latter is not included in our measurement of 
the torques). Gravitational torques are probably caused by the fact that both 
the halo and disk have asymmetries in their mass distribution which are caused 
in part by the initial noise in the halo models and in part arise as a result 
of evolution. In fact we find that gravitational torques are particularly strong 
where the disk develops non-axisymmetric features like spiral arms. 

\item Small losses of angular momentum by artificial viscous torques certainly occur, 
but at high resolution they yield the smallest contribution to hydrodynamical torques, 
which are dominated by those produced by spurious pressure gradients. These pressure 
torques could be associated with small lumps in the cold disk being dragged by the 
hot background medium.

\item Gas particles within the inner kpc transfer away nearly all of their initial
angular momentum. This is a physical transport that depends on the strength of the spiral patterns and the central bar but inappropriate  gravitational force softening can tamper those structures.

\end{itemize}

We found in our high-res runs that the gas particles which cool to the disk,  
lose overall $\sim20\%$ of their specific angular momentum while cooling as a 
result of the combination of the various effects mentioned.
The situation in cosmological simulations can only be worse since even fairly 
massive progenitors of a Milky Way-size
object have resolutions comparable to the lowest used in our work. Therefore numerical loss of angular 
momentum in the early stages of galaxy formation will be severe (larger than $50 \%$ 
based on our results). This could explain the excess of low angular momentum material always
present in simulated galaxies in the form of a slowly rotating, central bulge-like component; this forms 
typically only after a few billion years from the merging of two or more massive progenitors and is 
too concentrated and massive compared to bulges of late-type spirals (Governato et al. 2004).
This behaviour stands in contrast to several semi-analytical models which suppose 
strict angular momentum conservation while gas particles cool. 

One is tempted to conclude that in cosmological simulations a 
regime with negligible numerical loss will be hard to reach due to the nature of hierarchical structure
formation. Even very well resolved objects at the present day will host some baryonic component that was 
assembled earlier in poorly resolved objects. Some help comes from the fact that baryons would not
collapse in structures below $10^9 M_{\odot}$ or so due to the effect of reionisation -- if these are the 
first objects that we need to resolve well enough to have negligible artificial angular momentum loss
it means that the SPH particles mass has to be of order a few thousand solar masses, which implies
$> 10^7$ SPH particles will be required inside a final Milky-Way sized system. Although such a number
of SPH particles is still beyond current capabilities, improvement in both software and hardware 
will make this possible in the near future. It is interesting to note that a similarly high number of particles 
has been necessary in the past to completely control artifacts present in the dark matter component of 
simulations such as the overmerging of substructure (Ghigna et al. 2000).

In this work we also studied the morphology of the simulated galaxies. Bar formation should 
be a very common process for galaxies with masses comparable to the Milky Way: independent
of the mass resolution and of the use of the Katz-type star formation recipe, all the disks 
went bar unstable as soon as the force resolution reached $0.5$ kpc or $\sim0.25\%$ of $r_{200}.$  
Bars efficiently transport angular momentum towards the outer disk, which increases its scale length
by up to a factor of 2. This is an important effect to take into account in order to explain observed
galaxy sizes (see also Debattista et al. 2005). The formation of a strong bar in the runs with
star formation is clearly correlated with how quickly the disk attains a high enough mass.

Our results also show that not only the angular momentum but more  general the morphological evolution of simulated galaxies 
is affected by the mass and force resolution. The dependence of bar formation and the subsequent evolution of the
baryonic surface density profile on gravitational softening is a clear example of this. At this stage it is not clear what kind of 
convergence one should expect. In purely collisionless simulations, Weinberg \& Katz (2006) suggest that 
possibly billions of particles are needed to capture the gravitational interaction of a heavy bar and the halo, which controls
the angular momentum transport in the system. On the other hand Sellwood (2006) claims to be able to follow the slow down of the bar
due to the halo for a modest number of particles. The bars considered by Sellwood (2006) are weaker than those considered by
Weinberg \& Katz (2006), and are comparable to those in our models. Weinberg \& Katz (2006) point out that the difficulty
in treating the bar-halo interaction correctly lies in resolving the resonant structure of the combined 
stellar and dark matter potential. The same difficulty should in principle be present in our simulations, with the added complexity
introduced by the presence of a dissipational component. However, when a dissipative component is an important fraction of the mass 
in the system, if not completely dominant as in some of our models, the situation might be significantly different. It is hard to imagine 
that resonances would play a major role in this case, rather the large scale modes amplified by the self-gravity of the gaseous
disk (e.g. the spiral patterns or a gaseous bar)  would probably drive the evolution. Then the question becomes if and when the
growth of such global models converge. The nature of such modes will then affect star formation and this, in turn, the development
of the stellar disk (Li, Mac Low \& Klessen 2005). This is a major issue also in the field of star and planet formation. Convergence
tests starting from simple disk systems and with different codes represent the best way to pin down the problem and are under way 
(see http://www-theorie.physik.unizh.ch/$\sim$moore/wengen/tests). Preliminary results suggest that in this class of problems force resolution is the most critical aspect, since
it has to be small enough to capture the wavelength of the fastest growing modes already in the weakly nonlinear regime (Mayer et al., in preparation). Of course this adds to the other class of issues, namely how the nature of the cooling flow that forms the disk
changes with varying resolution (Kaufmann et al. 2006).

None of the runs, either Milky Way-like or M33-like, produced a purely exponential disk at high resolution.
In the runs with the Milky Way model a bar forms and with that a bimodal profile with a nearly exponential
outer disk and an inner bulge-like component with a much steeper profile. In the runs using an M33-like 
model a bar does not appear due to the low baryon fraction (and resulting much lighter disk) but a
power law-type profile arises which strongly deviates from exponential in the central region. The
steep central profile is associated with a dense nucleus. The nucleus is formed with particles that
lose most of their angular momentum, we believe as a result of transport by spiral arms. Although
the formation of the nucleus persisted even in a run where the mass resolution, thanks to particle
splitting, was ten times better than the standard hi-res runs, we cannot completely exclude that 
it is produced by a numerical artifact. 

The relationship between the non-exponential profiles seen in our work and the very similar power law
profiles predicted by semi-analytical models of galaxy formation (e.g. Van den Bosch 2001) is still to 
be understood. The comparison between the  semi-analytical models and the simulations is not straightforward, because the latter include self-gravity, and thus non-axisymmetric modes. Basic differences between
the semi-analytic models and simulations still exist. For example the gas does not cool onto the disk
in a spherical cooling flow, rather from a cylindrical region which extends above and below the disk.

It is tempting to associate the nucleus to the central dense stellar nuclei 
seen in several late-type spiral galaxies, i.e. in the limit of infinite resolution
then some transport of angular momentum remains which causes gas to quickly reach the central dark
matter cusp. Indeed the scale and corresponding overdensity of such a 
nucleus in the simulation is comparable to that seen, for example, in M33 (Regan \& Vogel 1994; Corbelli et al. 2003).  
M33 is not unique in this sense since similar nuclear components with a range of sizes and luminosities are found in many late-type spirals (Carollo et al. 1998, Carollo 1999). We verified that the surface density profiles and stellar nucleus are only slightly changed when we allow gas to be turned into stars. Our next step is to  extend the cooling function to metals and molecules
and include heating from supernovae winds and radiation backgrounds. One could 
imagine that feedback may cause a large fraction of the low angular momentum material of 
the inner nucleus to be removed, helping to produce a disk with exponential profile.

\section*{Acknowledgments}
We would like to thank Stelios Kazantzidis for providing a code to 
generate isolated dark matter halos. We acknowledge useful and stimulating discussions with Chiara Mastropietro, Frank van den Bosch, Aaron Dutton, Tom Quinn, Victor Debattista, Steen Hansen \& Fabio Governato. We  also thank the anonymous referee for valuable comments. The numerical simulations were performed on the 
zBox (http://krone.physik.unizh.ch/$\sim$stadel/zBox) supercomputer at the University of Z\"urich.

\label{lastpage}

\end{document}